\documentclass[11pt]{article}
\newcommand{\s}{\sigma}
\newcommand{\be}{\beta}
\newcommand{\ra}{\rightarrow}
\newcommand{\ua}{\uparrow}
\newcommand{\da}{\downarrow}

\newcommand{\mc}{\mathcal}
\newcommand{\el}{\ell}
\newcommand{\dd}{\delta}
\newcommand{\bra}{\langle}
\newcommand{\ket}{\rangle}
\newcommand{\mh}{\hat{m}}
\newcommand{\la}{\lambda}
\newcommand{\mr}{\mathrm}
\newcommand{\bhm}{\mbox{\boldmath$\hat{m}$}}
\newcommand{\bm}{\mbox{\boldmath$m$}}
\newcommand{\bmstar}{\mbox{\boldmath$m$}^{\star}}
\newcommand{\bhmstar}{\mbox{\boldmath$\hat{m}$}^{\star}}

\newcommand{\bxi}{\mbox{\boldmath$\xi$}}
\newcommand{\bxistar}{\mbox{\boldmath$\xi$}^{\star}}
\newcommand{\blambda}{\mbox{\boldmath$\lambda$}}
\newcommand{\blastar}{\mbox{\boldmath$\lambda$}^{\star}}

\newcommand{\bk}{\mbox{\boldmath$k$}}

\newcommand{\boldeta}{\mbox{\boldmath$\eta$}}

\newcommand{\bx}{\mbox{\boldmath$x$}}
\newcommand{\bs}{\mbox{\boldmath$\sigma$}}
\newcommand{\bsigma}{\mbox{\boldmath$\sigma$}}

\newcommand{\boldtau}{\mbox{\boldmath$\tau$}}

\begin{document}

\title{\Large\bf {\boldmath$1\!+\!\infty$} Dimensional Attractor Neural Networks}
\author{\bf N.S. Skantzos \hspace{10mm} A.C.C. Coolen\\[3mm]
Dept of Mathematics, 
King's College London, \\ The Strand, London WC2R 2LS,
UK \\
skantzos@mth.kcl.ac.uk \hspace{5mm} tcoolen@mth.kcl.ac.uk}
\maketitle

\begin{center}
PACS: 87.30, 05.20
\end{center}

\begin{abstract}
\noindent
We solve a class of attractor neural network models with a mixture
of 1D nearest-neighbour interactions and infinite-range
interactions, which are both of a Hebbian-type form. Our solution
is based on a combination of mean-field methods, transfer matrices,
and 1D random-field techniques, and is obtained both for
Boltzmann-type equilibrium (following sequential Glauber dynamics)
and Peretto-type equilibrium (following parallel dynamics).
Competition between the alignment forces mediated via short-range
interactions, and those mediated via infinite-range ones, is found
to generate novel phenomena, such as multiple locally stable `pure'
states, first-order transitions between recall states, 2-cycles and
non-recall states, and domain formation leading to extremely long
relaxation times. We test our results against numerical simulations
and simple benchmark cases, and find excellent agreement.
\end{abstract}

\tableofcontents

\clearpage
\section{Introduction}

Solvable models of recurrent neural networks are bound to be
simplified representations of biological reality. The early
statistical mechanical studies of such networks, e.g.\@
\cite{amitetal1,amitetal2}, concerned mean-field models, whose
statics and dynamics are by now well understood, and have obtained
the status of textbook material \cite{pnn}. The focus in
theoretical research has consequently turned to new areas, such as
solving the dynamics of large recurrent networks close to
saturation \cite{CLS}, the analysis of finite size phenomenology
\cite{CCV}, solving biologically more realistic models \cite{PNN2},
or networks with spatial structure
\cite{canning,noest1,domany,jonker,coolenviana}. In this paper we
analyse Ising spin models of recurrent networks with spatial
structure, in which there are two coexistent classes of
Hopfield-type \cite{amitetal1} interactions: infinite-range ones
(operating between any pair of neurons), and 1D short-range ones
(operating between nearest neighbours only). The study of this type
of structure is motivated by the interplay between long-range
processing via (excitatory) pyramidal neurons and short-range
processing via (inhibitory) inter-neurons, which is typically
observed in cortical tissue. In the present model however, and in
contrast to early papers on spatially structured networks, exact
solutions based solely on simple mean-field approaches are ruled
out. Due to short-range interactions analytical solutions require
significantly more complicated methods and the present models can
be solved exactly only by a combination of mean- and random-field
techniques \cite{brandtgross,bruinsma,aeppli,rujan,skantzos},
whereas for the special case in which the system has stored a
single pattern, a simple (Mattis) transformation allows us to
derive the solution via a combination of mean-field methods and
transfer-matrices.

Our paper is organised as follows. We first solve the one-pattern
case in which pattern-variables can be transformed away (thus
providing a convenient and exactly solvable benchmark case against
which to test the general theory). This also hints at the interesting
features induced by short- versus long-range competition in the
more general model. In particular, already in the one-pattern model
we find first-order phase transitions, regimes corresponding to
multiple locally stable states and we find that sequential and
parallel dynamics phase diagrams are related by simple
transformations. We then proceed to the general case, with an
arbitrary number of stored patterns, away from saturation regimes.
For sequential dynamics (Boltzmann equilibrium) we adapt the 1D
random-field techniques as originally developed for site-disordered
Ising chains, and combine them with mean-field methods; for
parallel dynamics (Peretto equilibrium) we adapt and combine with
mean field methods the procedure in \cite{skantzos}, based on
$4\times 4$ random transfer matrices. The disorder-averaged free
energy and the order parameters are found, in closed and exact
form, as integrals over the distribution of a characteristic
variable, which represents a specific ratio of conditioned
partition functions. This distribution is calculated following
\cite{brandtgross,bruinsma}.

In the region where the infinite-range versus short-range
competition is prominent, our theory predicts a
\emph{series} of (continuous and discontinuous) dynamic phase
transitions, and a free energy surface with multiple local minima.
These features become more prominent, in both number and strength,
when the number of stored patterns increases, in sharp contrast to
Hopfield-type infinite-range networks \cite{amitetal1}, where `pure
state' solutions are independent of the number of patterns stored.
The transition lines of sequential and parallel dynamics phase
diagrams are found to be related by reflection symmetries and also
parallel dynamics macroscopic equations can describe 2-cycles
rather than fixed-points solutions. Finally we test our theory
against several exactly solvable cases and against numerical
simulations; the latter are found to exhibit interesting but
extremely slow domain-induced dynamics, with plateau phases. Once
equilibration has occurred, we obtain excellent agreement between
theory and experiment.

\section{Model Definitions}

We study models with $N$ Ising spin  neuron variables
$\bs=(\s_1,\ldots,\s_{N})\in\{-1,1\}^N$, which evolve in time
stochastically on the basis of post-synaptic potentials $h_i(\bs)$
(or local fields), following the Glauber-type rule
\begin{equation}
{\rm Prob}[\sigma_i(t+1)=\pm 1]=\frac{1}{2}\left[1\pm \tanh[\beta
h_i(\bs(t))]\right]~~~~~~~~~~
h_i(\bs)=\sum_{j\neq i }J_{ij}\sigma_j+\theta_i
\label{eq:dynamics}
\end{equation}
The parameters $J_{ij}$ and $\theta_i$ represent synaptic
interactions and firing thresholds. The (non-negative) parameter
$\beta$ controls the amount of noise, with $\beta=0$ and
$\beta=\infty$ corresponding to purely random and purely
deterministic response, respectively. If the interaction matrix is
symmetric, both a random sequential execution and a fully parallel
execution of the stochastic dynamics (\ref{eq:dynamics}) will
evolve to a unique equilibrium state. The corresponding microscopic
state probabilities can both formally be written in the Boltzmann
form $p_\infty(\bs)\sim \exp[-\beta H(\bs)]$, with associated
Hamiltonians \cite{amitetal1,peretto} (since Peretto's
pseudo-Hamiltonian Hamiltonian $H_{\rm par}$ depends on $\beta$,
the associated statistics are not of the Boltzmann form):
\begin{eqnarray}
H_{\rm seq}(\bs)&=&-\frac12\sum_{i\neq j}\sigma_i
J_{ij}\sigma_j -\sum_i\theta_i\sigma_i
\\
H_{\rm par}(\bs)&=&
-\frac{1}{\beta}\sum_{i}\log 2\cosh[\beta
h_i(\bs)] -\sum_i\theta_i\sigma_i
\label{eq:peretto}
\end{eqnarray}
In both cases, expectation values of order parameters  can be
obtained by differentiation of the free energy per neuron
$f=-\lim_{N\to\infty}(\beta N)^{-1}\log
\sum_{\bs}\exp[-\beta H(\bs)]$, which acts as a
generating function. For the parameters $J_{ij}$ and $\theta_i$ we now make the following choice
\footnote{Competition between a different type of uniform infinite-range and
random nearest-neighbour interactions has been studied recently in
\cite{vieira}, for sequential dynamics only.}:
\begin{equation}
J_{ij}=\frac{J^{\el}_{ij}}{N}
+J^{s}_{ij}\,(\delta_{i,j+1}+\delta_{i,j-1})
\hspace{15mm}
\theta_i=\theta
\label{eq:model_2}
\end{equation}
\[
J^{\ell}_{ij}=J_{\ell}^{(1)}+
J_\ell^{(2)}\sum_{\mu=1}^{p}\xi_{i}^{\mu}\xi_{j}^{\mu}
\hspace{15mm}
J^{s}_{ij}=J_{s}^{(1)}+
J_{s}^{(2)}\sum_{\mu=1}^{p}\xi_{i}^{\mu}\xi_{j}^{\mu}
\]
This corresponds to the result of having stored a set of binary
patterns $\{\bxi^1,\ldots,\bxi^p\}$ with $\xi_i^\mu\in\{-1,1\}$ in
a one-dimensional chain of neurons,  through Hebbian-type learning,
but with different (potentially conflicting) embedding strengths
$J_s^{(2)}$ and $J_\ell^{(2)}$ associated with the short-range
versus the infinite-range interactions. We will choose $p\ll N$.
The parameters $J_s^{(1)}$ and $J_\ell^{(1)}$ control uniform
contributions to the interactions within their class. Taking
derivatives of $f$ with respect to $J_{\ell}^{(2)}$ produces
expressions involving the familiar `overlap' order parameters:
\begin{eqnarray*}
{\rm sequential:} & &
\bm^{2}=-2\,\frac{\partial f}{\partial J_{\ell}^{(2)}}
=\lim_{N\ra\infty}
\bra(\frac{1}{N}\!
\sum_{i}\s_{i}\bxi_{i})^{2}\ket_{\mr{eq}}
\\
{\rm parallel:} & &
\bm^2=-\frac{\partial f}{\partial
J_{\el}^{(2)}}=\lim_{N\ra\infty}\bra(\frac{1}{N}\!
\sum_{i}\s_{i}\bxi_{i})^2\ket_{\mr{eq}}
\end{eqnarray*}
where the brackets $\bra\ldots\ket_{\mr{eq}}$ denote equilibrium
averages, $\bm=(m_1,\ldots,m_p)$, and
$\bx_i\cdot\bx_j=\sum_{\mu}x_i^\mu\xi_j^\mu$. Note that for
$J_{s}^{(2)}=J_{\ell}^{(2)}=0$ we obtain the simpler model
\begin{equation}
J_{ij}=\frac{J_{\el}}{N}+J_{s}(\delta_{i,j+1}+\delta_{i,j-1})
~~~~~~~~~~~~~~
\theta_i=\theta
\label{eq:model_Ib}
\end{equation}
The Mattis transformation $\s_i\to\s_i \xi_i$ maps this model onto
\begin{equation}
J_{ij}=\frac{J_{\el}}{N}\xi_i\xi_j+J_{s}(\delta_{i,j+1}+\delta_{i,j-1})\xi_i\xi_j
~~~~~~~~~~~~~~
\theta_i=\theta\xi_i
\label{eq:model_Ia}
\end{equation}
which corresponds to the result of having stored just a single
pattern $\bxi=(\xi_1,\ldots,\xi_N)\in\{-1,1\}^N$. Taking
derivatives of $f$ with respect to the parameters $\theta$ and
$J_s$ in (\ref{eq:model_Ib}) produces our order parameters:
\begin{eqnarray*}
\mr{seq:}
& &
m=-\frac{\partial f}{\partial\theta}=\lim_{N\to\infty}
\frac{1}{N}\sum_i \bra \sigma_i\ket_{\mr{eq}}
\hspace{7mm}
a=-\frac{\partial f}{\partial J_s}=\lim_{N\to\infty}
\frac{1}{N}\sum_i\bra \sigma_{i+1}\sigma_{i}\ket_{\mr{eq}}
\\
\mr{par:}
& &
m=-\frac{1}{2}\frac{\partial f}{\partial\theta}=\lim_{N\to\infty}
\frac{1}{N}\sum_{i} \bra \sigma_{i}\ket_{\mr{eq}}
\hspace{7mm}
a=-\frac{1}{2}\frac{\partial f}{\partial J_s}=\lim_{N\to\infty}
\frac{1}{N}\sum_i\bra\sigma_{i+1}\tanh[\beta
h_i(\bs)]\ket_{\mr{eq}}
\end{eqnarray*}
where we have simplified the parallel dynamics observables
with the identities
\[
\bra \sigma_{i+1}\tanh[\beta h_i(\bs)]\ket_{\mr{eq}}=
\bra \sigma_{i-1}\tanh[\beta h_i(\bs)]\ket_{\mr{eq}}
\hspace{10mm}
\mr{and}
\hspace{10mm}
\bra \tanh[\beta h_i(\bs)]\ket_{\mr{eq}}=\bra
\sigma_i\ket_{\mr{eq}}
\]
which follow from (\ref{eq:dynamics}) and from invariance under the
transformation $i\to N+ 1- i$ (for all $i$). For model
(\ref{eq:model_Ib}) $m$ is the average neuronal activity. For
sequential dynamics $a$ describes the average equilibrium state
covariances of neighbouring neurons, and for parallel dynamics it
gives the average equilibrium state covariances of neurons {\em at
a given time} $t$, and their neighbours {\em at time} $t+1$ (the
difference between the two meanings of $a$ will be important in the
presence of 2-cycles). For model (\ref{eq:model_Ia}) one similarly
finds
\[
m=\lim_{N\to\infty}\frac{1}{N}\sum_i \bra\xi_i
\sigma_i\ket_{\mr{eq}}
\hspace{15mm}
a=\lim_{N\ra\infty}\frac{1}{N}\sum_{i}
\bra\,(\xi_{i}\sigma_{i})(\xi_{i+1}\s_{i+1})\,\ket_{\mr{eq}}
\]
The observable $m$ is here the familiar overlap order parameter of
associative memory models \cite{amitetal1,amitetal2}, which
measures the quality of pattern recall in equilibrium. Note that
$m,a\in[-1,1]$.

%

\section{Solution and Phase Diagrams for $p=1$\label{sec:sol_modelI}}

Before we proceed to the solution of the general model
(\ref{eq:model_2}) we first solve the relatively simple
situation, where a single pattern has been stored following
(\ref{eq:model_Ib}) and (\ref{eq:model_Ia}). This has the following
advantages: it being a simpler version of (\ref{eq:model_2}), it
allows us to explore general features and build intuition,  without
as yet any serious technical subtleties, as brought up by the
general model. At the same time it provides an excellent benchmark
test of the general theory to which it should reduce for
$J_{s}^{(2)}=J_{\ell}^{(2)}=0$. For the remainder of this section
our analysis will refer to model (\ref{eq:model_Ib}).

\subsection*{Solution via Transfer Matrices}

In calculating the asymptotic free energy per neuron $f$ it is
advantageous to separate terms induced by the long-range synapses
from those induced by the short-range ones, via insertion of
$1=\int\!dm~\delta[m-\frac{1}{N}\sum_i \sigma_i]$. Upon using the
integral representation of the $\delta$-function, we then arrive at
\[
f=-\lim_{N\to\infty}\frac{1}{\beta N}\log
\int\!dmd\mh~e^{-\beta N\phi(m,\mh)}
\vspace*{-3mm}
\]
with
\begin{eqnarray*}
\phi_{\rm seq}(m,\mh)&=&-im\mh -m\theta -\frac{1}{2}J_\el m^2
-\frac{1}{\beta N}\log R_{\rm seq}(\mh)
\\
\phi_{\rm par}(m,\mh)&=&-im\mh-m\theta
-\frac{1}{\beta N}\log R_{\rm par}(m,\mh)
\end{eqnarray*}
The quantities $R$ contain all complexities due to the short-range
interactions in the model. They correspond to
\begin{eqnarray*}
R_{\rm seq}(\mh)&=&\sum_{\bs\in\{-1,1\}^{N}}
e^{-i\be \mh\sum_{i}\s_{i} }
\ e^{\beta J_s\sum_i \s_i\s_{i+1}}
\\
R_{\rm par}(m,\mh)&=&\sum_{\bs\in\{-1,1\}^{N}}
e^{-i\be \mh\sum_{i}\s_{i}}\
\prod_{i}\log [2\cosh [\be J_\el m+ \be \theta+
\be J_s(\sigma_{i+1}+ \sigma_{i-1})]]
\end{eqnarray*}
They can be calculated using the transfer-matrix method giving
\[
R_{\rm seq}(\mh)=\mathrm{Tr}~[\textbf{\emph{T}}^{N}_{\rm seq}]
~~~~~~~~~~~~~~~~~~~~~
\textbf{\emph{T}}_{\rm seq}=\left(\begin{array}{cc}
e^{\be J_{s}-i\beta \mh} & e^{-\be J_{s}} \\[1mm]
e^{-\be J_{s}} & e^{\be J_{s}+i\beta \mh}\end{array}\right)
~~~~~~~~~~~~
\]
\[
R_{\rm par}(m,\mh)=\mathrm{Tr}~[\textbf{\emph{T}}^{N}_{\rm par}]
~~~~~~~~~~~~~
\textbf{\emph{T}}_{\rm par}=\left(
\begin{array}{cc}
2\cosh[\be w_{+}]\, e^{-i\beta \mh} & 2\cosh[\be w_0]
\\[1mm]
2\cosh[\be w_0] & 2\cosh[\be w_{-}]\, e^{i\beta \mh}
\end{array}\right)
\]
where $w_0= J_\el m+\theta$ and $w_{\pm}= w_0\pm 2J_{s}$.
The identity
$\mathrm{Tr}~[\textbf{\emph{T}}^{N}]
=\lambda_+^N+\lambda_-^N$, in
which $\lambda_\pm$ are the eigenvalues of the $2\times 2$ matrix
$\textbf{\emph{T}}$ enables  us to take the limit $N\to\infty$ in
our equations. The integral over $(m,\mh)$ is then for $N\to\infty$
evaluated by steepest descent, and is dominated by the saddle
points of the exponent $\phi$. We thus arrive at the transparent
result
\begin{equation}
f=\mathrm{extr}~\phi(m,\mh)~~~~~~~~~~
\left\{
\begin{array}{l}
\phi_{\rm seq}(m,\mh)=-im\mh-m\theta -\frac{1}{2}J_\el m^2
-\frac{1}{\beta}\log\la_{+}^{\rm seq}\\[2mm]
\phi_{\rm par}(m,\mh)=-im\mh-m\theta
-\frac{1}{\beta}\log \la_{+}^{\rm par}
\end{array}
\right.
\label{eq:modeI_phis}
\end{equation}
where $\la_{+}^{\rm seq}$ and $\la_{+}^{\rm par}$ are the largest
eigenvalues of $\textbf{\emph{T}}_{\rm seq}$ and
$\textbf{\emph{T}}_{\rm par}$:
\begin{eqnarray*}
\lambda_{+}^{\mr{seq}}&=&
e^{\be J_{s}}\cosh[i\be\hat{m}]+\left[e^{2\be J_{s}}\cosh^2[i\be
\hat{m}]-2\sinh[2\be J_{s}]\right]^{\frac12}
\\
\lambda_{+}^{\mr{par}}&=&
\cosh[\be w_+]e^{-i\be \hat{m}}+\cosh[\be w_-]e^{i\be \hat{m}}+
\left[\cosh^2[\be w_{+}]e^{-2\be i \hat{m}}+\right.
\\
& &
\hspace{20mm}
\left.+\cosh^2[\be w_-]e^{2\be i\hat{m}}-
2\cosh[\be w_+]\cosh[\be w_-]+4\cosh^2[\be w_0]\right]^{\frac12}
\end{eqnarray*}
For simplicity, we will restrict ourselves to the case where
$\theta=0$; generalisation of what follows to the case of arbitrary
$\theta$, by using the full form of (\ref{eq:modeI_phis}), is not
significantly more difficult. The expressions defining the value(s)
of the order parameter $m$ can now be obtained from the saddle
point equations
$\partial_m\phi(m,\mh)=\partial_{\mh}\phi(m,\mh)=0$. This is a
straightforward differentiation task for the sequential case. For
parallel, one obtains a set of coupled non-linear equations, namely
\begin{equation}
m=\mc{F}(m,\tilde{m}) \hspace{10mm} \tilde{m}=\mc{F}(\tilde{m},m)
\label{eq:saddle_parallel}
\end{equation}
where $\tilde{m}=-i\hat{m}/J_{\ell}$ and $\mc{F}(\cdot\,,\cdot)$ corresponds to
\[
\mc{F}(p,q)=\Delta(p,q)^{-1}\left(e^{2\be J_{s}}\sinh[\be J_{\ell}(p+q)]
-e^{-2\be J_{s}}\sinh[\be J_{\ell}(p-q)]\right)
\]
\[
\Delta(p,q)=\left[e^{2\be J_{s}} \sinh^2[\be J_{\ell}(p+q)]+e^{-2 \be J_s}
\sinh^2[\be J_{\ell}(p-q)] +2\cosh^2[\beta J_{\ell}p]+2\cosh^2[\beta
J_{\ell}q]\right]^\frac12
\]
with $\Delta(p,q)=\Delta(q,p)$. We will now show that the parallel
dynamics fixed point problem (\ref{eq:saddle_parallel}) admits the
unique solution $m=\mathrm{sgn}[J_{\ell}]\,
\tilde{m}$, by the following argument:
\[
\begin{array}{lcc}
{\rm For}\ J_{\ell}\geq 0:  m=\tilde{m}
& {\rm since} & 0\leq(m-\tilde{m})^2=
\Omega(m,\tilde{m})\ (m-\tilde{m})\,\sinh[\be J_{\ell}(\tilde{m}-m)]\,\leq
0 \\
{\rm For}\ J_{\ell}< 0: m=-\tilde{m}
& {\rm since} & 0\leq(m+\tilde{m})^2=
\Omega(m,\tilde{m})\ (m+\tilde{m})\,\sinh[\be J_{\ell}(\tilde{m}+m)]\,\leq 0
\end{array}
\]
\[
{\rm where}
\hspace{10mm}
\Omega(m,\tilde{m})\equiv 2\,e^{-2\beta J_{s}} \Delta(m,\tilde{m})^{-1}>0
\]
Insertion of these solutions to the original function
$\mc{F}(\cdot\,,\cdot)$ allows us to reduce
(\ref{eq:saddle_parallel}) to a simple 1D fixed-point problem,
similar in structure with what follows from the sequential case,
namely
\begin{equation}
\begin{array}{lllllll}
{\rm sequential:} && \mh=i m J_\ell, &&
m=G(m;J_\ell,J_s)
\\[3mm]
{\rm parallel:} && \mh=im J_\ell, &&
m=G(m;J_\ell,J_s) && {\rm for}~~J_\ell\geq 0 \\[1mm]
 && \mh=-im J_\ell, &&
 m=G(m;-J_\ell,-J_s) && {\rm for}~~J_\ell< 0
\label{eq:saddle}
\end{array}
\end{equation}
with
\begin{equation}
G(m;J_\ell,J_s)=\frac{\sinh[\beta J_\ell m]}{\sqrt{\sinh^2[\beta
J_\ell m]+e^{-4\beta J_s}}}
\label{eq:map}
\end{equation}
The macroscopic
observable $a$ is generated by differentiating the
reduced free energy per neuron (\ref{eq:modeI_phis}):
\begin{equation}
\begin{array}{lllllll}
{\rm sequential:} && \mh=i m J_\ell, &&
a=F(m;J_\ell,J_s)
\\[3mm]
{\rm parallel:} && \mh=im J_\ell, &&
a=F(m;J_\ell,J_s) && {\rm for}~~J_\ell\geq 0 \\[1mm]
 && \mh=-im J_\ell, &&
 a=F(m;-J_\ell,-J_s) && {\rm for}~~J_\ell< 0
\end{array}
\label{eq:a}
\end{equation}
with
\begin{equation}
F(m;J_\ell,J_s)=\frac{\cosh[\be J_{\el}m]\sqrt{\sinh^2[\be J_{\ell}m]+e^{-4\be
J_{s}}}+\sinh^{2}[\be J_{\ell} m]-e^{-4\be J_{s}}}
{\cosh[\be J_{\el}m]\sqrt{\sinh^2[\be J_{\ell}m]+e^{-4\be
J_{s}}}+\sinh^{2}[\be J_{\ell} m]+e^{-4\be J_{s}}}
\label{eq:a_map}
\end{equation}
Note that in the absence of short-range interactions we recover the
familiar Curie-Weiss law $m=\tanh[\be J_{\ell}m]$ whereas in the
absence of long-range interactions (\ref{eq:saddle}) and
(\ref{eq:a}) reduce to $m=0$ and $a=\tanh[\beta J_s]$, as they
should. Finally, it is worth noting that equations
(\ref{eq:saddle},\ref{eq:map},\ref{eq:a},\ref{eq:a_map}) allow us
to derive the physical properties of the parallel dynamics model
from those of the sequential dynamics model via simple parameter
transformations.

\subsection*{Phase Transitions \& Phase Diagrams}

Our main  order parameter  $m$ is to be solved from an equation of
the form $m\!=\!G(m)$, in which $G(m)\!=\!G(m;J_\el,J_s)$ for both
sequential and parallel dynamics with $J_\ell\!\geq \!0$, whereas
$G(m)\!=\!G(m;-J_\ell,-J_s)$ for parallel dynamics with $J_\ell<0$.
Note that, due to $G(0;J_\el,J_s)\!=\!0$, the trivial solution
$m\!=\!0$ always exists. In order to obtain a phase diagram we have
to perform a bifurcation analysis of the equations
(\ref{eq:saddle},\ref{eq:map}), and determine the combinations of
parameter values for which specific non-zero solutions are created
or annihilated (the transition lines). Bifurcations of non-zero
solutions occur when simultaneously $m=G(m)$ (saddle-point
requirement) and $1=\partial_m G(m)$ ($m$ is in the process of
being created/annihilated). Analytical expressions for the lines in
the $(\beta J_s,\beta J_\el)$ plane where second-order transitions
occur between recall states (where $m\!\neq \!0$)  and non-recall
states (where $m\!=\!0$) are obtained by solving the coupled
equations $m=G(m)$ and $1=\partial_m G(m)$ for $m=0$. This gives:
\begin{equation}
{\mathit cont.~trans.:}~~~~~~~~~~
\begin{array}{ll}
\mathrm{sequential:}\  & \be J_{\el}=e^{-2\be J_{s}} \\ [2mm]
\mathrm{parallel:}\ &  \be J_{\el}=e^{-2\be J_{s}}\hspace{4mm}
 \mathrm{and}
\hspace{4mm} \be J_{\el}=-e^{2\be J_{s}}
\end{array}
\label{eq:second_order}
\end{equation}
whereas for the macroscopic observable $a$ we obtain
\[
\mr{sequential/parallel:}
\hspace{15mm}
a=\tanh[\be J_s]
\]
If along the lines (\ref{eq:second_order}) we inspect the behaviour
of $G(m)$ close to $m\!=\!0$ we can anticipate the possible
existence of first-order transitions, using the properties of
$G(m)$ for $m\to\pm \infty$, in combination with $G(-m)=- G(m)$.
Precisely {\em at} the lines (\ref{eq:second_order}) we have
$G(m)=m+\frac{1}{6}G^{'''}(0).m^3+\mc{O}(m^5)$. Since
$\lim_{m\ra\infty}G(m)=1$ one knows that, when
$G^{\prime\prime\prime}(0)\!>\!0$,  a discontinuous transition must
have already taken place earlier, and that away from the lines
(\ref{eq:second_order}) there will consequently be regions where
one finds five solutions of $m\!=\!G(m)$ (two positive ones, two
negative ones and $m=0$). Along the lines (\ref{eq:second_order})
the condition $G^{\prime\prime\prime}(0)>0$ translates into
\begin{equation}
\begin{array}{ll}
\mathrm{sequential:}\ & \ \, \be J_{\el}>\sqrt{3}
\hspace{3mm}\mathrm{and}\hspace{4mm}\be
J_{s}<-\frac{1}{4}\log{3} \\ [2mm]
\mathrm{parallel:}\ & |\be J_{\el}|>\sqrt{3}
\hspace{3mm}\mathrm{and}\hspace{3mm}|\be
J_{s}|<-\frac{1}{4}\log{3}
\end{array}
\label{eq:nomore_cont}
\end{equation}
In the present models it turns out that one can also find an
analytical expression for the discontinuous transition lines in the
$(\beta J_s,\beta J_\el)$ plane, in the form of a parametrisation.
For sequential dynamics one finds a single line, parametrised by
$x=\be J_{\el}m\in[0,\infty)$:
\begin{equation}
{\mathit discont.~trans.:}
~~~~~~~~
\be J_{\el}(x)=\sqrt{\frac{x^{3}}{x-\tanh(x)}},
~~~~~~~~
 \be J_{s}(x)=-\frac{1}{4}\log
\left[\frac{\tanh(x)\sinh^{2}(x)}{x-\tanh(x)}\right]
\label{eq:first_order}
\end{equation}
This can be verified by explicit substitution into
(\ref{eq:saddle}). Since this parametrisation
(\ref{eq:first_order}) obeys $\beta J_s(0)=-\frac{1}{4}\log 3$  and
$\beta J_\el(0)=\sqrt{3}$, the discontinuous transition indeed
starts precisely at the point predicted by the convexity of  $G(m)$
at $m=0$, see (\ref{eq:nomore_cont}). In the limit $x\to\infty$ the
slope of (\ref{eq:first_order}) approaches $\beta
J_{\ell}(\infty)/\beta J_s(\infty)\to -2$. For sequential dynamics
the line (\ref{eq:first_order}) gives all non-zero solutions of the
bifurcation requirements $m=G(m)$ and $1=\partial_m G(m)$. For
parallel dynamics one finds, in addition to (\ref{eq:first_order}),
a second `mirror' transition line, generated  by the transformation
$\{\beta J_{\ell}, \beta J_{s}\}\mapsto
\{-\beta J_{\ell}, -\beta J_{s}\}$.

\begin{figure}[th]
\vspace*{61mm}
\hbox to
\hsize{\hspace*{-1mm}\includegraphics{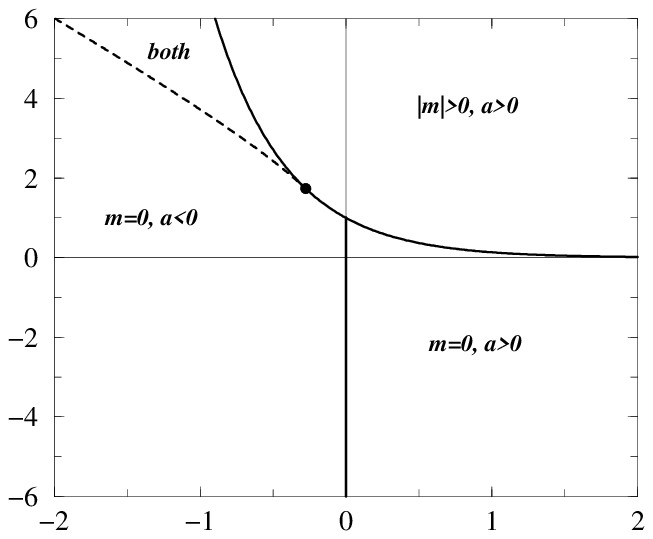}\hspace*{1mm}}
\vspace*{-5mm}
\hbox to
\hsize{\hspace*{75mm}\includegraphics{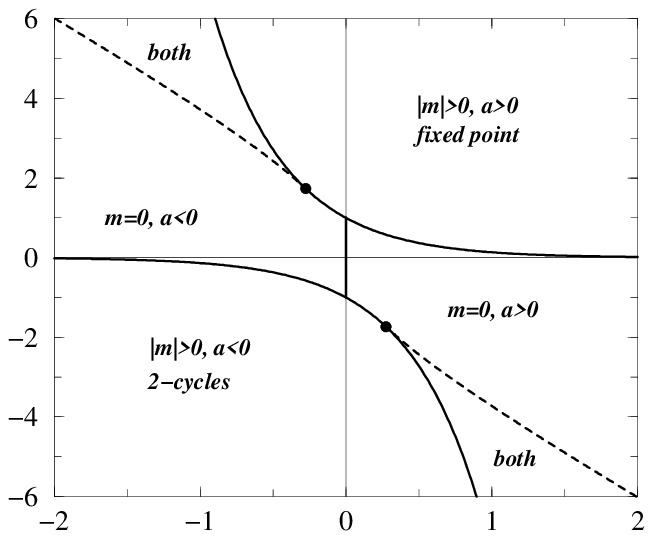}\hspace*{-75mm}}
\vspace*{-32mm}
\begin{picture}(200,85)(10,50)
\hspace*{-5mm}
{\Large
\put(20,145){ $\beta J_{\ell}$}
\put(234,145){ $\beta J_{\ell}$}
\put(140,55){ $\beta J_{s}$}
\put(350,55){ $\beta J_{s}$}
}
\end{picture}
\vspace*{-2mm}
\caption{\small
Left: Phase diagram for sequential dynamics, involving: (\emph{i})
a region with $m=0$ only (here $a=\tanh[\beta J_s]$), (\emph{ii}) a
region where both the $m=0$ state and two $m\neq 0$ states are
locally stable, and (\emph{iii}) a region with two locally stable
$m\neq 0$ states (with opposite sign, and with identical $a>0$).
All solid lines indicate second-order transitions, whereas the
dashed lines indicate first-order ones.
 Right: phase diagram for parallel dynamics, involving the
above regions and transitions, as well as a second set of
transition lines (in the region $J_\el<0$) which are exact
reflections in the origin of the first set. Here, however, the two
$m\neq 0$ physical solutions describe 2-cycles rather than
fixed-points, and the $J_\el<0$ region describes simultaneous local stability
of the $m=0$ fixed-point and 2-cycles.}
\label{fig:diagrams}
\end{figure}

Having determined the transition lines in parameter space, we can
turn to the phase diagrams. Figure \ref{fig:diagrams} shows the
phase diagram for the two types of dynamics, in the $(\beta
J_s,\beta J_\el)$ plane (note: of the three parameters
$\{\beta,J_s,J_\ell\}$ one is redundant). For sequential dynamics
we find ($i$) a region where $m=0$ only ($ii$) a region where the
trivial as well as two non-trivial states (a positive and a
negative one) are all locally stable (selection will thus be based
on initial conditions) and ($iii$) a region where only $m\neq 0$
states are localy stable. The transitions ($i)\to(iii$) and
($ii)\to (iii$) are second-order ones (solid lines of figure
\ref{fig:diagrams}) whereas the transition ($i)\to(ii$) is
first-order (dashed line). In region ($i$), where $m=0$ only, one
finds $a=\tanh[\beta J_s]$ whereas in regions $(ii)$ and $(iii)$
$a$ is given by the full expression of (\ref{eq:a}). For parallel
dynamics we find, in addition to the sequential phase transitions,
a second set of `mirror' transition lines generated by $\{\beta
J_\ell, \beta J_s\}\mapsto\{-\beta J_\ell, -\beta J_s\}$. In
contrast to the sequential case however, here in the region $\beta
J_{\ell}<0$ and $\beta J_s<0$ where $m\neq 0$ can be a physical
state (lower left corner of the phase diagram of figure
\ref{fig:diagrams}) one finds 2-cycles between the two $m\neq 0$
(positive and negative) recall states. This can be inferred from
the exact dynamical solution that is available along the line
$J_s=0$ (see e.g.\@ \cite{pnn}), given by the deterministic map
$m(t+ 1)=\tanh[\beta J_\el m(t)]$. This map gives a stable period-2
oscillation for $\beta J_\el<- 1$, of the form $m(t)=(-1)^t
m^\star$, where $m^\star=\tanh[\beta |J_\el|m^\star]$. In
 the 2-cycle region one has
$a=\lim_{N\to\infty}\frac{1}{N}
\sum_i \bra \sigma_{i+1}\tanh[\beta h_i(\bsigma)]\ket<0$.
This can be understood on the basis of the (parallel dynamics) identity
$\bra \sigma_{i+1}\tanh[\beta h_i(\bsigma)]\ket=\bra
\sigma_{i+1}(t)\sigma_i(t+ 1)\ket$.

\begin{figure}[t]
\vspace*{61mm}
\hbox to
\hsize{\hspace*{3mm}\includegraphics{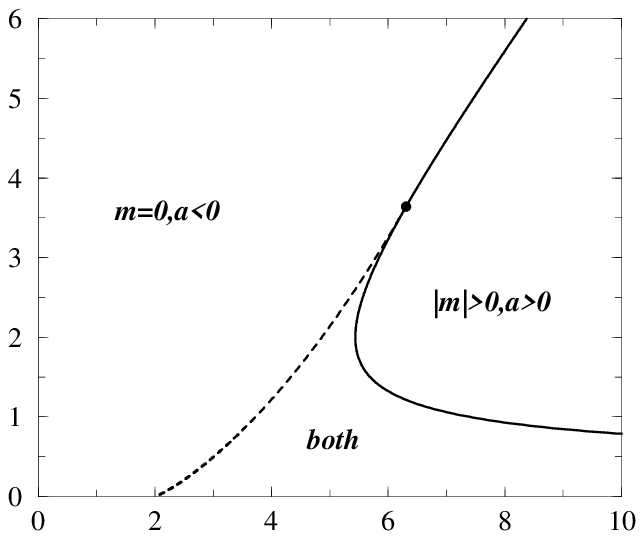}\hspace*{-3mm}}
\vspace*{-4mm}
\hbox to
\hsize{\hspace*{78mm}\includegraphics{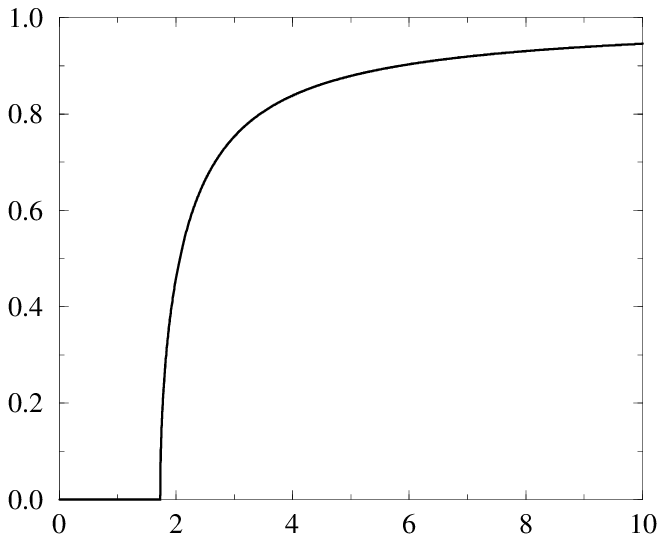}\hspace*{-78mm}}
\vspace*{-32mm}
\begin{picture}(200,85)(10,50)
\hspace*{-5mm}
{\Large
\put(35,150){ $T$}
\put(250,155){ $m$}
\put(160,50){ $J_{\ell}$}
\put(365,50){ $\beta J_\ell$}
}
\end{picture}
\vspace*{-1mm}
\caption{\small Left: alternative presentation
of the competition region of the sequential dynamics phase diagram
of figure \ref{fig:diagrams}. Here the system states  and
transitions are drawn in the $(J_\el,T)$ plane ($T=\beta^{-1}$),
for $J_{s}=-1$. Right: the magnitude of the `jump' of the overlap
$m$ along the first-order transition line (dashed lines in figure
\ref{fig:diagrams}), as a function of $\beta J_\el(x)$,
$x\in[0,\infty)$, see equation (\ref{eq:first_order}).}
\label{fig:alternative}
\end{figure}

We find that in contrast to models with nearest neighbour
interactions only ($J_\ell=0$, where no pattern recall will occur),
and to models with mean-field interactions only ($J_s=0$, where
pattern recall can occur), the combination of the two interaction
types leads to qualitatively new modes of operation, especially in
the competition region, where $J_\ell>0$ and $J_s<0$ (Hebbian
long-range synapses, combined with anti-Hebbian short range ones).
The novel features of the diagram can play a useful role: the
existence of multiple locally stable states ensures that only
sufficiently strong recall cues will evoke pattern recognition; the
discontinuity of the transition subsequently ensures that in the
latter case the recall will be of a substantial quality. In the
case of parallel, similar statements can be made in the opposite
region of synaptic competition, but now involving 2-cycles. Since
figure \ref{fig:diagrams} cannot show the zero noise region
$(\beta=T^{-1}=\infty)$, we have also drawn the interesting
competition region of the sequential dynamics phase diagram in the
$(J_\el,T)$ plane, for $J_s=-1$ (see figure \ref{fig:alternative},
left picture). At $T=0$ one finds coexistence of recall states
$(m\neq 0)$ and non-recall states $(m=0)$ for any $J_\el>0$, as
soon as $J_s<0$. In the same figure (right picture) we show the
magnitude of the discontinuity in the order parameter $m$ along the
discontinuous transition line, as a function of $\beta J_\el$.

\begin{figure}[h]
\vspace*{65mm}
\hbox to
\hsize{\hspace*{3mm}\includegraphics{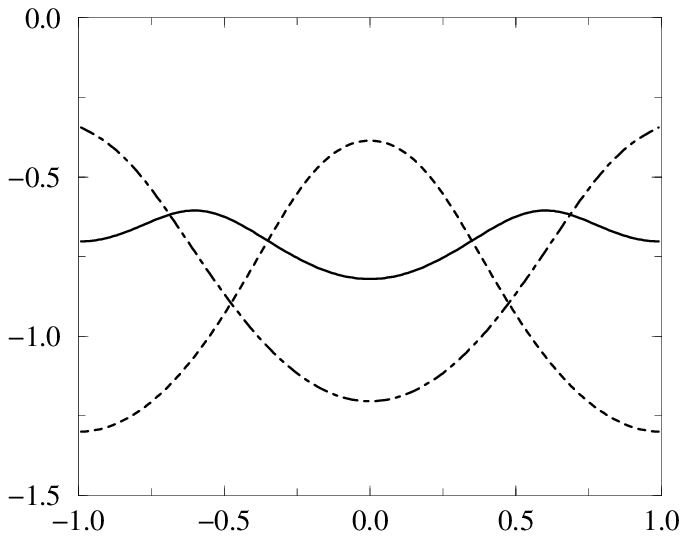}\hspace*{-3mm}}
\vspace*{-4.5mm}
\hbox to
\hsize{\hspace*{78mm}\includegraphics{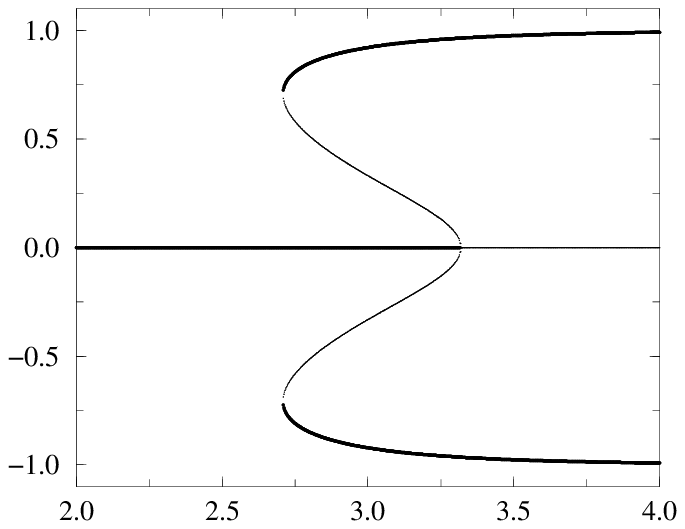}\hspace*{-78mm}}
\vspace*{-32mm}
\begin{picture}(200,85)(10,50)
\hspace*{-5mm}
{\Large
\put(25,145){ $\phi(m)$}
\put(250,145){ $m$}
\put(150,50){ $m$}
\put(360,50){ $\beta J_\ell$}
}
\end{picture}
\vspace*{-1mm}
\caption{\small Left: Free energy per neuron $\phi(m)=
\phi_{\mathrm {seq}}(m,im J_{\ell})$
as derived from equation (\ref{eq:modeI_phis}) for
$(\beta,J_{\ell})=(2,3)$. The three lines correspond to regimes
where ($i$) $m\neq0$ only (dashed line: $J_s=-0.2$) ($ii$) trivial
and non-trivial states are both locally stable (solid line:
$J_s=-0.8$), and ($iii$) $m=0$ only (dot-dashed line: $J_s=-1.2$).
Right: Sequential dynamics bifurcation diagram displaying for $\be
J_s=-0.6$ the possible recall solutions. For small $\beta J_\ell$,
$m=0$ is the only stable state. At a critical $\beta J_\ell$ given
by (\ref{eq:first_order}), $m$ jumps to non-zero values. For
increasing $\beta J_{\ell}$ the unstable $m\neq0$ solutions (thin
lines) converge towards the trivial one until $\beta
J_{\ell}=\exp(-\beta J_s)$ (here: $\beta J_{\ell}\approx 3.32$)
where a second-order transition takes place and $m=0$ becomes
unstable.}
\label{fig:energy}
\end{figure}

Finally we show, by way of further illustration of the coexistence
mechanism, the value of reduced exponent $\phi_{\mathrm{seq}}(m)$
given in (\ref{eq:modeI_phis}), evaluated upon elimination of the
auxiliary order parameter $\mh$: $\phi(m)\!\equiv\! \phi_{\rm
seq}(m,imJ_\el)$. The result, for the parameter choice
$(\beta,J_{\ell})\!=\!(2,3)$ and for three different short-range
coupling stengths (corresponding to the three phase regimes:
non-zero recall, coexistence and zero recall) is given in figure
\ref{fig:energy}. In the same figure we also give the sequential
dynamics bifurcation diagram displaying the value(s) of the overlap
$m$ as a function of $\beta J_\ell$ and for $\beta J_{s}\!=\!-0.6$
(a line crossing all three phase regimes in figure
\ref{fig:diagrams}).

\section{Solution and Phase Diagrams for Arbitrary $p$}

In the general case (\ref{eq:model_2}), where $p>1$, the pattern
variables $\{\xi_i^{\mu}\}$ can no longer be transformed away as in
model (\ref{eq:model_Ia}); here in order to arrive at expressions
for the order parameters one has to perform the disorder average of
the free energy over the realisation of $\{\xi_{i}^{\mu}\}$. Due to
the addition of nearest neigbour interactions, however, the
disorder average has become significantly more complicated than
that in infinite-range models, even in the loading regime away from
saturation (i.e.\@ $\lim_{N\to\infty}p/N=0$). We perform the
disorder average via a suitable adaptation of 1D random-field
techniques, see e.g.\@ \cite{brandtgross,bruinsma,skantzos}. These
are based on the derivation of a stochastic process in which
observables of systems of size $N$ are mapped to observables of
systems of size $N+1$. We will assume that in the thermodynamic
limit this process leads to a unique stationary state. The free
energy is then given as an integral over the distribution of a
characteristic ratio of conditioned partition functions, and the
order parameters follow via differentiation.

\subsection*{Adaptation of RFIM Techniques}

We introduce the notation $\bmstar=(m_0,\bm)$,
$\bhmstar=(\hat{m}_0,\bhm)$, $\bxistar=(\xi^0,\bxi)$ where
$\xi^0=1$. We separate the overlap order parameter via insertion of
$1\!=\!\int\! d\bmstar\
\dd[\bmstar-1/N\sum_{i}\s_{i}\bxistar_{i}]$, and we replace the
$\dd$-functions by their integral representation giving
\[
f=-\lim_{N\ra\infty}\frac{1}{\be N}\log \int\!d\bmstar
d\bhmstar\ e^{-\be N \phi_{N}(\bmstar,\bhmstar)}
\]
where
\begin{eqnarray}
\phi_{N}^{\mr{seq}}(\bmstar,\bhmstar)&=&-i\bmstar\cdot\bhmstar
-m_0\,\theta-\frac{1}{2}J_{\ell}^{(1)}m_0^2-\frac12
J_{\ell}^{(2)}\bm^2-
\frac{1}{\be N}\,\log R^{\mr{seq}}_N (\bhmstar)
\label{eq:phis_seq}
\\
\phi_{N}^{\mr{par}}(\bmstar,\bhmstar)
&=&-i \bmstar\cdot\bhmstar-m_0\,\theta-\frac{1}{\be
N}\,\log R^{\mr{par}}_N(\bmstar,\bhmstar)  \label{eq:phis_par}
\end{eqnarray}
The quantities $R_N(\bmstar,\bhmstar)$ contain, as in model
(\ref{eq:model_Ib}) in section \ref{sec:sol_modelI}, the summation
over $\bs=(\s_1,\ldots,\s_N)$ and the short-range neuron
interactions
\begin{equation}
R^{\mr{seq}}_N(\bhm)=\sum_{\bs\in\{-1,1\}^N}
\mc{F}_{\rm seq}(\bs)
\hspace{15mm}
R^{\mr{par}}_N(\bmstar,\bhmstar)=\sum_{\bs\in\{-1,1\}^N}\mc{F}_{\rm par}(\bs),
\label{eq:Rseq}
\end{equation}
\begin{eqnarray*}
\mc{F}_{\rm seq}(\bs) &=&
e^{-i\be \sum_{i}\s_{i}\bhmstar\cdot\bxistar_{i}}
\ e^{\be\sum_{i}\s_{i}J^{s}_{i,i+1}\s_{i+1}}
\\
\mc{F}_{\rm par}(\bs)&=&
e^{-i\,\be\sum_i
 \s_i \,\bhmstar\cdot\bxistar_{i}}\
\prod_{i}\log[2\cosh
[\be (J_{i}^{\el}+\s_{i-1}J^{s}_{i-1,i}
+\s_{i+1}J^{s}_{i,i+1})]]
\end{eqnarray*}
where the short-hands $J^{s}_{i,i+1}$ and $J_{i}^{\el}$
correspond to:
\[
J_{i,i+1}^{s}=J_{s}^{(1)}+J_{s}^{(2)}\,
\bxi_{i}\cdot\bxi_{i+1}
\hspace{15mm}
J^{\el}_{i}=\theta+J_{\el}^{(1)}m_0+J_{\el}^{(2)}
\bxi_i\cdot\bm
\]
Each of the quantities in (\ref{eq:Rseq}) is  now separated into
different constituent parts, defined by conditioning on either the
state of the last neuron in the chain (sequential dynamics), or on
the states of the last two neurons in the chain (parallel
dynamics):
\begin{equation}
\left[\begin{array}{c}
R^{\mr{seq}}_{N,\ua} \\[2mm]
R^{\mr{seq}}_{N,\da}
\end{array}\right]
=
\sum_{\bs}
\mc{F}_{\rm seq}(\bs)
 \left[\begin{array}{l}
\dd_{\s_{N},1}  \\[2mm]
\dd_{\s_{N},-1}
\end{array}\right]
\label{eq:define_Rseq}
\end{equation}

\begin{equation}
\left[\begin{array}{c}
R^{\mr{par}}_{N,\ua\ua} \\[2mm]
R^{\mr{par}}_{N,\ua\da} \\[2mm]
R^{\mr{par}}_{N,\da\ua} \\[2mm]
R^{\mr{par}}_{N,\da\da}
\end{array}\right]
=
\sum_{\bs} \mc{F}_{\rm par}(\bs)
  \left[\begin{array}{l}
\dd_{\s_{N-1},1}\ \dd_{\s_{N},1} \\ [2mm]
\dd_{\s_{N-1},1}\ \dd_{\s_{N},-1} \\ [2mm]
\dd_{\s_{N-1},-1}\ \dd_{\s_{N},1} \\ [2mm]
\dd_{\s_{N-1},-1}\ \dd_{\s_{N},-1}
\end{array}\right]
\label{eq:define_Rpar}
\end{equation}
We now add an extra neuron to the chain. After some simple
bookkeeping and assuming non-periodic boundary conditions we derive
the following recurrence relations:
\begin{equation}
{\rm seq:}\hspace{5mm}
\left(\begin{array}{l}
R_{N+1,\ua} \\ R_{N+1,\da}
\end{array}\right)=
\left(\begin{array}{ll}
e^{\beta(J^{s}_{N,N+1}-i{ \bhmstar\cdot\bxistar_{N+1}})} &
e^{-\beta(J^{s}_{N,N+1}+i{ \bhmstar\cdot\bxistar_{N+1}})} \\
e^{-\beta(J^{s}_{N,N+1}-i{ \bhmstar\cdot\bxistar_{N+1}})}  &
e^{\beta(J^{s}_{N,N+1}+i{ \bhmstar\cdot\bxistar_{N+1}})}
\end{array}\right)
\left(\begin{array}{l}
R_{N,\ua} \\ R_{N,\da}
\end{array}\right)  \label{eq:seqmatrix}
\end{equation}

\begin{eqnarray}
{\rm par:}  \!&\! \!&\!
\left(\!\begin{array}{l}
R_{N+1,\ua\ua} \\ R_{N+1,\ua\da}
\end{array}\!\right)=
2\,Q_{N,N+1}^{+}
\left(\!\begin{array}{cc}
  \frac{e^{-i\beta{ \bhmstar\cdot\bxistar_{N+1}}}}
         {Q^{+}_{N-1,N}}A^{(N)}_{(+,+)}
  &  \frac{e^{-i\beta{ \bhmstar\cdot\bxistar_{N+1}}}}{Q^{-}_{N-1,N}}
  A^{(N)}_{(-,+)}\\[3mm]
  \frac{e^{i\beta { \bhmstar\cdot\bxistar_{N+1}}}}{Q^{+}_{N-1,N}}A^{(N)}_{(+,-)}
  & \frac{e^{i\beta { \bhmstar\cdot\bxistar_{N+1}}}}{Q^{-}_{N-1,N}} A^{(N)}_{(-,-)}
\end{array}\!\!\right)\!
\left(\!\begin{array}{l}
R_{N,\ua\ua} \\ R_{N,\da\ua}
\end{array}\!\right) \nonumber
\\ [4mm]
 \!&\! \! &\!
\left(\!
\begin{array}{l}
R_{N+1,\da\ua} \\ R_{N+1,\da\da}
\end{array}\!\right)=
2\,Q_{N,N+1}^{-}
\left(\!\begin{array}{cc}
\frac{e^{-i\beta{ \bhmstar\cdot\bxistar_{N+1}}}}{Q^{+}_{N-1,N}}A^{(N)}_{(+,+)}
&  \frac{e^{-i\beta\bhmstar\cdot\bxistar_{N+1}}}{Q^{-}_{N-1,N}}
A^{(N)}_{(-,+)}\\[3mm]
\frac{e^{i\beta\bhmstar\cdot\bxistar_{N+1}}}{Q^{+}_{N-1,N}}A^{(N)}_{(+,-)}
& \frac{e^{i\beta\bhmstar\cdot\bxistar_{N+1}}}{Q^{-}_{N-1,N}} A^{(N)}_{(-,-)}
\end{array}\!\!\right)\!
\left(\!\begin{array}{l}
R_{N,\ua\da} \\ R_{N,\da\da}
\end{array}\!\right)\label{eq:parmatrix}
\end{eqnarray}
with the short-hand notation:
\[
A^{(n)}_{(\pm,\pm)}
=
A^{(n)}_{(\pm,\pm)}(
\bxi_{n-1}\cdot\bxi_{n}\,,\,
\bxi_{n}\cdot\bxi_{n+1}\,,\,\bxi_{n}\cdot\bm)
= \cosh[\,\be(J_{n}^{\ell}\,\pm\,
J^s_{n-1,n}\,\pm\, J^{s}_{n,n+1})]
\]
\begin{equation}
Q_{n,n+1}^{\pm}=
Q^{\pm}(\bxi_{n}
\cdot\bxi_{n+1}\,,\,\bxi_{n}\cdot\bm)=
\cosh[\be(J^{\el}_{n+1}\pm J^{s}_{n,n+1})]
\label{eq:shorthand}
\end{equation}
The fact that parallel dynamics leads to a recurrence process given
by $4\times 4$ random matrices, instead of the simpler $2\times 2$
random matrices of the sequential case, is due to the appearance of
short-range couplings of the form
$cosh[\beta(J^{\ell}_i+\s_{i-1}J^{s}_{i-1,i}+
\s_{i+1}J^s_{i,i+1})]$ in (\ref{eq:Rseq}), rather than
the more familiar exponentials. This, in turn, is an immediate
consequence of the form of the pseudo-Hamiltonian
(\ref{eq:peretto}). We also note that in the parallel case the two
reduced $2\times 2$ matrices differ only by the prefactors
$Q_{N,N+1}^{\pm}$. The recurrence matrices
(\ref{eq:seqmatrix}-\ref{eq:parmatrix}) will form the basis for
evaluating the free energy per neuron, which follows from
\begin{equation}
R_{N}^{\rm seq}=R_{N,\ua}^{\rm seq}+R_{N,\da}^{\rm seq}
\label{eq:conditionedR-seq}
\end{equation}
\begin{equation}
R_{N}^{\rm par}=R_{N,\ua\ua}^{\rm par}+R_{N,\ua\da}^{\rm par}+
R_{N,\da\ua}^{\rm par}+R_{N,\da\da}^{\rm par}
\label{eq:conditionedR-par}
\end{equation}

\subsection*{The Stochastic Process for Conditioned Partition Functions}

In the spirit of \cite{brandtgross,bruinsma}
we next define the ratios between the conditioned quantities
(\ref{eq:define_Rseq}) and (\ref{eq:define_Rpar}), and study their
stochastic `evolution' generated by adding new neurons
successively to the chain:
\begin{equation}
\hspace{-60mm}
\mr{sequential:}
\hspace{10mm}
 k_{n+1}=e^{-2\be i\bhmstar\cdot\bxistar_{n+1}}\
\frac{R_{n+1,\da}}{R_{n+1,\ua}}
\label{eq:define_kseq}
\end{equation}
\[
\mr{parallel:}
\hspace{10mm}
 k_{n+1}^{(1)}=e^{-2\be i \bhmstar\cdot\bxistar_{n+1}}
\frac{R_{n+1,\ua\da}}{R_{n+1,\ua\ua}}
\hspace{10mm}
k_{n+1}^{(2)}=e^{-2\be i \bhmstar\cdot\bxistar_{n+1}}
\frac{R_{n+1,\da\da}}{R_{n+1,\da\ua}}
\]
\begin{equation}
\hspace{-43mm}
k_{n+1}^{(3)}=
\frac{Q_{n,n+1}^{+}}{Q_{n,n+1}^{-}}\
\frac{R_{n+1,\da\da}}{R_{n+1,\ua\da}}
\label{eq:define_kpar}
\end{equation}
Due to their dependence on the random variables $\{\bxi\}$, each of
these quantities is a stochastic variable. Insertion of
(\ref{eq:seqmatrix}-\ref{eq:parmatrix}) into the above relations gives
\begin{eqnarray}
\mr{sequential:} & &
k_{n+1}=
\frac{e^{-\be J^{s}_{n,n+1}}+k_{n} \ e^{2\be i\bhmstar\cdot\bxistar_{n}}
\ e^{\be J^{s}_{n,n+1}}}
{e^{\be J^{s}_{n,n+1}}+k_{n} \ e^{2\be i\bhmstar\cdot\bxistar_{n}}\ e^{-\be
J^{s}_{n,n+1}}} \label{eq:kseq}
\\ [3mm]
\mr{parallel:} & &
k_{n+1}^{(1)}=\frac{ A_{(+,-)}^{(n)}\
k_{n}^{(2)}+ A_{(-,-)}^{(n)}\
k_{n}^{(1)}\,k_{n}^{(3)}}
{ A_{(+,+)}^{(n)}\
k_{n}^{(2)}+ A_{(-,+)}^{(n)}\
k_{n}^{(1)}\,k_{n}^{(3)}}
\hspace{12mm}
k_{n+1}^{(2)}=\frac{ A_{(+,-)}^{(n)}
+ A_{(-,-)}^{(n)}\ k_{n}^{(3)}}
{A_{(+,+)}^{(n)}+ A_{(-,+)}^{(n)}\
k_{n}^{(3)}} \nonumber
\\
& &
k_{n+1}^{(3)}=\frac{ A_{(+,-)}^{(n)}
\ k_{n}^{(1)}\,k_{n}^{(2)}+ A_{(-,-)}^{(n)}\
k_{n}^{(1)}\,k_{n}^{(2)}\,k_{n}^{(3)}}
{ A_{(+,-)}^{(n)}\
k_{n}^{(2)}+ A_{(-,-)}^{(n)}\
k_{n}^{(1)}\,k_{n}^{(3)}}\
e^{2\be i\bhmstar\cdot\bxistar_{n}}
\label{eq:kpar}
\end{eqnarray}
For parallel dynamics we observe that, if the identity
$k^{(1)}_{n}=k^{(2)}_{n}$ is true, then also
$k^{(1)}_{n+1}=k^{(2)}_{n+1}$. Furthermore, it can be easily
checked that this is the case for $n=2$. We are thus guaranteed
that $k^{(1)}_{n}=k^{(2)}_{n}$ for all $n\geq 2$, and now all three
quantities $\{k^{(1)}_{i},k^{(2)}_{i},k^{(3)}_{i}\}$ can (for any
$i=2,\ldots,N$) be expressed in terms of a single stochastic
variable which we will take to be $k^{(1)}_{i}\equiv k_{i}$. This
simplifies considerably the description of the parallel dynamics
stochastic process:
\[
k_{n+1}^{(1)}=k_{n+1}
\hspace{10mm}
k_{n+1}^{(2)}=\frac{ A_{(+,-)}^{(n)}
+ A_{(-,-)}^{(n)}\ k_{n-1}\, e^{2\beta i\bhmstar\cdot\bxistar_{n-1}} }
{A_{(+,+)}^{(n)}+ A_{(-,+)}^{(n)}\
k_{n-1}\,e^{2\beta i\bhmstar\cdot\bxistar_{n-1}}}
\hspace{10mm}
k_{n+1}^{(3)}=k_{n}\,e^{2\beta i\bhmstar\cdot\bxistar_{n}}
\]
We have thus obtained  1D stochastic maps for both sequential and
parallel dynamics:
\begin{eqnarray}
\mr{seq:}\hspace{4mm}
k_{i+1}&=&\psi_{\mr{seq}}\left
(k_{j}\,;\,\bxi_{j}\cdot\bxi_{j+1}
\,,\,\bhmstar\cdot\bxistar_{j}\,|\,\forall\, j\leq i\right)
\label{eq:seqmap}
\\
&=&
\frac{e^{-\be J^{s}_{n,n+1}}+k_{n} \ e^{2\be i\bhmstar\cdot\bxistar_{n}}
\ e^{\be
J^{s}_{n,n+1}}}
{e^{\be J^{s}_{n,n+1}}+k_{n} \ e^{2\be i\bhmstar\cdot\bxistar_{n}}\ e^{-\be
J^{s}_{n,n+1}}}    \nonumber
\\
\mr{par:}\hspace{4mm}
k_{i+1}&=&\psi_{\mr{par}}\left(
k_{j-1}\,;\,\bxi_{j-1}\cdot\bxi_{j}\,,\,\bxi_{j}\cdot
\bxi_{j+1}\,,\,\bm\cdot\bxi_{j}\,,\,
\bhmstar\cdot\bxistar_{j-1}\,|\,\forall\, j\leq i\right)  \label{eq:parmap}
\\
& =&
\frac{ \cosh[\beta(J^{\el}_i+J^s_{i-1,i}-J^s_{i,i+1})] +
       e^{2\beta i\bhmstar\cdot\bxistar_{i-1}}k_{i-1}
       \cosh[\beta(J^{\el}_i-J^s_{i-1,i}-J^s_{i,i+1})] }
{ \cosh[\beta(J^{\el}_i+J^s_{i-1,i}+J^s_{i,i+1})] +
      e^{2\beta i\bhmstar\cdot\bxistar_{i-1}}k_{i-1}
       \cosh[\beta(J^{\el}_i-J^s_{i-1,i}+J^s_{i,i+1})] }
 \nonumber
\end{eqnarray}
We observe that, in contrast to the sequential dynamics mapping,
parallel dynamics distinguishes between even and odd sites, and
produces two independent sets of stochastic variables $\{k_j\}$.

\subsection*{Disorder Averaging and the Free Energy}

One can now express the non-trivial part of the free energy in
terms of the above ratio's $k_i$, using
(\ref{eq:conditionedR-seq}-\ref{eq:conditionedR-par}):
\[
-\frac{1}{\be N}\log R_{N}^{\rm seq}=-\frac{1}{\be
N}\log R_{N,\ua}^{\rm seq}+\mc{O}(\frac1N)
\hspace{10mm}
-\frac{1}{\be N}\log R_{N}^{\rm par}=-\frac{1}{\be
N}\log R_{N,\ua\ua}^{\rm par}+\mc{O}(\frac1N)
\]
Upon also using (\ref{eq:define_kseq}-\ref{eq:define_kpar}) and
(\ref{eq:kseq}-\ref{eq:kpar}) iteratively, in order to map the
quantities $\{R_{n,\ua}^{\rm seq},R_{n,\ua\ua}^{\rm par}\}$ onto
$\{R_{n-1,\ua}^{\rm seq},R_{n-1,\ua\ua}^{\rm par}\}$ (for all
$n\leq N$), one can write the above expressions in the form:
\begin{eqnarray*}
-\frac{1}{\be N}\log R_{N,\ua}^{\mr{seq}}(\bmstar)
\!\!&\!\!=\!\!&\!\!
-\frac{1}{\be N}
\sum_{i}\log \left[e^{\beta J^{s}_{i,i+1}}+
e^{-\beta J^{s}_{i,i+1}}\ e^{2\be i \bhmstar\cdot\bxistar_{i}}k_{i}
\right]
\\
& & \hspace{3mm}
+\frac1N\sum_{i}i\bhmstar\cdot\bxistar_{i+1}
+\mc{O}(\frac1N)
\\
& &
\\
-\frac{1}{\be N}\log R_{N,\ua\ua}^{\mr{par}}(\bmstar,\bhmstar)
\!\!&\!\!=\!\!&\!\!
-\frac{1}{\be N}\sum_{i}\log\left[
   2\cosh[\beta(J^{\ell}_{i+1}+J^{s}_{i,i+1}+J^{s}_{i+1,i+2})]
    +  \right.
\\
& &
  \hspace{30mm} \left. +2\cosh[\beta(J^{\ell}_{i+1}-J^{s}_{i,i+1}+J^{s}_{i+1,i+2})]
   \, e^{2\beta i \bhmstar\cdot\bxistar_{i}}\,k_{i} \right]
\\
& &
+\frac{1}{N}\sum_{i}i\bhmstar\cdot\bxistar_{i+1}
-\frac{1}{\be N}\sum_{i}\log\left[Q^{+}_{i+1,i+2}/
Q^{-}_{i+1,i+2}\right]
+\mc{O}(\frac1N)
\end{eqnarray*}
where the terms $Q^\pm_{j,k}$ are defined at (\ref{eq:shorthand}). For
$N\ra\infty$ the above expressions are self-averaging and the
boundary terms vanish, giving

\[
\hspace{-83mm}
\mr{sequential:}
\hspace{8mm}
-\lim_{N\ra\infty}\frac{1}{\be N}
\log R_{N}^{\mr{seq}}(\bhmstar)=
\]
\begin{equation}
=
i\bra\bhmstar\cdot\bxistar\ket_{\bxi}
-\frac{1}{\be}\sum_{\bxi'\bxi}\int dk'\ P^{\mr s}(k',\bxi',\bxi)
\ \log\left[e^{\beta (J_s^{(1)}+J_s^{(2)}\bxi'\cdot\bxi)}+e^{\beta (J_s^{(1)}+J_s^{(2)}\bxi'\cdot\bxi)}
\,e^{2\be i\bhmstar\cdot{\bxistar}'}k'\right]
\label{eq:nontrivial_seq}
\end{equation}

\[
\hspace{-77mm}
\mr{parallel:}
\hspace{11mm}
-\lim_{N\ra\infty}\frac{1}{\be
N}\log R_{N}^{\mr{par}}(\bmstar,\bhmstar)=
\]
\[
\hspace{-35mm}
=i\bra \bhmstar\cdot{\bxistar}'\ket_{\bxi'}-\frac{1}{\be}\log 2
-\frac{1}{\be}\,\bra\,\log\left[\frac{Q^{+}(\bxi'\cdot\bxi,
\bxi\cdot\bm)}
{Q^{-}(\bxi'\cdot\bxi,
\bxi\cdot\bm)}\right]\,\ket_{\{\bxi',\bxi\}}-
\]
\begin{equation}
-\frac{1}{\be}\!\!\sum_{\bxi''\bxi'\bxi}\!\!\int\!\! dk''\,
 P^{\mr p}\!(k''\!,\bxi''\!,\bxi'\!,\bxi)
\log\!\left[\!A_{(+,+)}(\bxi''\!\cdot\!\bxi',
\bxi'\!\cdot\!\bxi,\bxi'\!\cdot\!\bm)
+
A_{(-,+)}(\bxi''\!\cdot\!\bxi',
\bxi'\!\cdot\!\bxi,\bxi'\!\cdot\!\bm)
\,k''\,e^{2\beta i \bhmstar\cdot{\bxistar}^{\prime\prime}}
\right]  \label{eq:nontrivial_par}
\end{equation}
in which the probability distributions $P^{\mr s}(\ldots)$ and
$P^{\mr p} (\ldots)$ are  defined as follows:
\begin{eqnarray}
\mr{sequential:}\hspace{4mm} P^{\mr s}(k',\bxi',\bxi)& =&
\lim_{N\ra\infty}\frac{1}{N}\sum_{i}\,\bra
\dd[k'-k_{i}]\,\,\dd[\bxi'-\bxi_{i}]
\,\dd[\bxi-\bxi_{i+1}]\,\ket_{\mr {eq}} \nonumber
\\
&= &
\frac{1}{2^{p}}\ P^{\mr s}(k',\bxi')
\label{eq:factor_seq}
\\
\mr{parallel:} \hspace{4mm}P^{\mr p}(\bk'',\bxi'',\bxi',\bxi) &= &
\lim_{N\ra\infty}\frac{1}{N}\sum_{i}\,\bra
\, \dd[\bk''-\bk_{i-1}]\,
\dd[\bxi''-\bxi_{i-1}]\,\dd[\bxi'-\bxi_{i}]\,\dd[\bxi-\bxi_{i+1}]
\,\ket_{\mr {eq}} \nonumber
\\
&=&
\frac{1}{2^{2p}}\ P^{\mr p}(\bk'',\bxi'')
\label{eq:factor_par}
\end{eqnarray}
In order to arrive at factorisation of these joint probability distributions we
have used the fact that, due to the form of the stochastic maps
 (\ref{eq:seqmap}) and (\ref{eq:parmap}), the quantities
$\{k_{n}\}$ are (for both types of dynamics) independent of
$\{\bxi_{n+1}\}$, i.e.\@ of the pattern variables at the next site.
The dependence on $\{\bxi_n\}$, however, does not allow us to write
down immediately (\ref{eq:seqmap}-\ref{eq:parmap}) in terms of a
Markov process. We thus introduce the auxiliary variables
$\{\blambda_n\}$, so that $\{\bxi_n\}$ itself becomes one of the
stochastically evolving variables, and the extended process becomes
indeed Markovian:
\[
\hspace{-22mm}
\mathrm{seq:}
\hspace{5mm}
\left(\!\begin{array}{l}
k \\ \blambda
\end{array}\!\right)_{i+1}
=
\Psi_{\mr{seq}}\left[\left(\!
\begin{array}{l}
k \\ \blambda
\end{array}\!\right)_{i};
\bxi_{i+1}\right]
=
\left(\!\!\begin{array}{c}
\psi_{\mr{seq}}(k_{i};\blambda_{i}\cdot\bxi_{i+1},
\bhmstar\cdot\blastar_{i}) \\ \bxi_{i+1}
\end{array}\!\!\right)
\]
\[
\mathrm{\mr par:}
\hspace{2mm}
\left(\!\begin{array}{l}
k \\ \blambda
\end{array}\!\right)_{i+1}
=
\Psi_{\mr {par}}\left[\left(
\!\begin{array}{l}
k \\ \blambda
\end{array}\!\right)_{i-1};
\bxi_{i+1},\bxi_{i}\right]
=
\left(\!\!\begin{array}{c}
\psi_{\mr{par}}
(k_{i-1};\blambda_{i-1}\cdot\bxi_{i},
\bxi_{i}\cdot\bxi_{i+1},\bm\cdot\bxi_{i},
\bhmstar\cdot\blastar_{i-1}) \\ \bxi_{i+1}
\end{array}\!\!\right)
\]
Equivalently:
\begin{equation}
\hspace{-10mm}
\mr{sequential:}
\hspace{10mm}
P_{i+1}^{\mr {s}}(k,\blambda)=\frac{1}{2^p}
   \sum_{\blambda'}\int dk'\ W_{\mr{s}}\left[
   \left(\begin{array}{l}
    k' \\ \blambda'\end{array}\right)\ra
   \left(\begin{array}{l}
   k \\ \blambda\end{array}\right)\right]\ P_{i}^{\mr {s}}(k',\blambda')
   \label{eq:density_seq}
\end{equation}
\begin{equation}
\mr{parallel:}
\hspace{10mm}
P_{i+1}^{\mr{p}}(k,\blambda)=
   \frac{1}{2^{p}}\sum_{\blambda''}\int dk''\
   W_{\mr{p}}\left[
   \left(\begin{array}{l}
   k'' \\ \blambda'' \end{array}\right)\ra
   \left(\begin{array}{l}
   k \\ \blambda \end{array}\right)\right]\
   P_{i}^{\mr{p}}(k'',\blambda'')
\label{eq:density_par}
\end{equation}
where for the transition probabilities $W_{\mr{s}}[\ldots]$ and
$W_{\mr{p}}[\ldots]$ we define:
\[
\hspace{-22mm}
\mr{sequential:}
\hspace{7mm}
W_{\mr{s}}\left[\left(\begin{array}{l}
k' \\ \blambda'\end{array}\right)
\ra
\left(\begin{array}{l}
k \\ \blambda\end{array}\right)\right]=
\bra\,\dd_{\bxi,\blambda}\,
\dd[k-\psi_{\mr{seq}}(k',\blambda'
\cdot\bxi,\bhmstar\cdot{\blastar}')]\
\,\ket_{\bxi}
\]
\[
\hspace{-0mm}
\mr{parallel:}
\hspace{5mm}
W_{\mr{p}}
\left[\left(\begin{array}{l}
k'' \\ \blambda'' \end{array}\right)
\ra
\left(\begin{array}{l}
k \\ \blambda \end{array}\right)\right]
=
\bra\,\dd_{\bxi,\blambda}\,
   \dd[k-\psi_{\mr{par}}(k'',
   {\blambda}''\cdot\bxi',\,\bxi'
   \cdot\bxi,\bm\cdot\bxi',\bhmstar\cdot{\blastar}'')]\
   \,\ket_{\bxi\bxi'}
\]
The evaluation of the non-trivial part of the free energies
(\ref{eq:nontrivial_seq},\ref{eq:nontrivial_par}) has now been
reduced to determining the (stationary) distributions of
(\ref{eq:density_seq}-\ref{eq:density_par}), which will be denoted
as $P_\infty^{\rm s}(k)$ and $P_{\infty}^{\rm p}(k)$. To achieve
this final objective we follow \cite{brandtgross,bruinsma} and
introduce the integrated densities $\hat{P}(k)=\int_{0}^{k}\, dz\,
P(z)$. After some algebra to eliminate the $\delta$-functions of
the transition probabilities via  the identity $\int\!
dx\,\dd[g(x)]f(x)=f\left(g^{inv}(0)\right)/|\partial_x g(0)|$, we
then obtain the following recursive relation:
\begin{eqnarray}
\mr{sequential:}\hspace{6mm}
\hat{P}_{i+1}^{\mr{s}}(k,\blambda)
\!&= &\!
\frac{1}{2^p}\sum_{\blambda'}\hat{P}_{i}^{\mr{s}}
\left(B_{\mr{s}}(k\,;\,\blambda'\cdot\blambda\,,\,\bhmstar\cdot{\blastar}'),
\blambda'\right)  \label{eq:integrated_seq}
\\
\mr{parallel:}\hspace{6mm}
\hat{P}_{i+1}^{\mr{p}}(k,\blambda)
\!& =&\!
\frac{1}{2^{2p}}\!\sum_{\blambda''}\sum_{\bxi'}\!
    \hat{P}_{i}^{\mr{p}}\left(B_{\mr{p}}(k\,;\,{\blambda}''\cdot\bxi'\,,\,
    \bxi'\cdot\blambda\,,\bm\cdot\bxi'\,,\,\bhmstar\cdot{\blastar}'),
    \blambda''\right)  \label{eq:integrated_par}
\end{eqnarray}
where the functions $B_{\mr{s}}(k)$ and
$B_{\rm p}(k)$ correspond to
\[
B_{\mr{s}}(k)= \frac{k\,e^{\be J_{s}}-
e^{-\be J_{s}}} {e^{\be
J_{s}}-
k\,e^{-\be J_{s}}}\, e^{-2\be i
\bhmstar\cdot{\blastar}'}
\hspace{10mm}
B_{p}(k)=
  \frac{k\,A_{(+,+)}-A_{(+,-)}}
       {A_{(-,-)}-kA_{(-,+)}}\,
e^{-2\be i\bhmstar\cdot{\blastar}''}
\]
For notation simplicity we have removed the arguments of $J_s$ and
$A_{(\pm,\pm)}$. These correspond to
\[
\begin{array}{lcl}
J_{s}\!&\!=\!&\!J_s(\blambda'\cdot\blambda)
\\[2mm]
\!&\!=\!&\!J_{s}^{(1)}+J_{s}^{(2)}\blambda'\cdot\blambda
\end{array}
\hspace{12mm}
\begin{array}{lcl}
A_{(\pm,\pm)}\!&\!=\!&\!A_{(\pm,\pm)}
({\blambda}''\cdot\bxi',\bxi'\cdot\blambda,\bm\cdot\bxi')
\\ [2mm]
\!&\!=\!&\!\cosh[\be(J^{\el}(\bm\cdot\bxi',m_0,\theta)\pm
J_s(\blambda''\cdot\bxi')
\pm J_s(\bxi'\cdot\blambda))]
\end{array}
\]
We have thus derived as yet fully exact expressions for the
disorder-averaged free energies
(\ref{eq:nontrivial_seq},\ref{eq:nontrivial_par}), as integrals
over the distribution of the stochastic quantities $\{k_{\rm
seq},k_{\rm par}\}$, which can be evaluated by numerical iteration
of the recursive relation
(\ref{eq:integrated_seq}-\ref{eq:integrated_par}).

\subsection*{The `Pure State' Ansatz and Symmetries of the Model}

We will now look for the so-called `pure-state' solutions: we will
assume that for $N\to\infty$ there is only one component $m_{\mu}$
of order $\mc{O}(1)$  (for simplicity we will take $\mu=1$) whereas
for all $\mu>1$  $m_{\mu}$ is of order $\mc{O}(N^{-\frac12})$:
$\bmstar=(m_{0},m,0,\ldots,0)$ and
$\bhmstar=(\hat{m}_0,\hat{m},0,\ldots,0)$. This is the standard
ansatz made in (infinite-range) associative memory models, which
gives the dominant states of the system. The fact that this is also
true for the present type of models is supported by numerical
simulations, see section \ref{sec:simulations}. Using this
simplification we will now prove that the stationary integrated
densities $\hat{P}_{\infty}^{\mr{s}}(\ldots)$ and
$\hat{P}_{\infty}^{\mr{p}}(\ldots)$ are independent of all
non-condensed pattern components. This will be shown by induction.
We will show that if $\hat{P}_i^{\rm s}(k,\blambda)$ and
$\hat{P}_i^{\rm p}(k,\blambda)$ are independent of $\la_{\mu}$ for
some $\mu>1$, then $\hat{P}_{i+1}^{\rm s}(k,\blambda)$ and
$\hat{P}_{i+1}^{\rm p}(k,\blambda)$ will also be independent of
$\lambda_\mu$. This will then immediately imply that if the
densities are independent of $\{\la_{\mu}\}$ for all $\mu>1$ at
step $i$ of the process
(\ref{eq:integrated_seq}-\ref{eq:integrated_par}), this will remain
so for all $\mu>1$ at any step $j>i$. Usage of the spin-flip
operator $F_{\mu}\bx=(x_1,\ldots,-x_{\mu},\ldots,x_{p})$ allows us
to write
\[
\begin{array}{lcl}
\mr{seq:}\hspace{6mm}
\Delta_{i+1}^{\rm s}\!\!&\!\!=\!\!&\!\!\hat{P}_{i+1}^{\mr{s}}
(k,F_{\mu}\blambda)-\hat{P}_{i+1}^{\mr{s}}(k,\blambda)
\\[2mm]
\!\!&\!\!=\!\!&\!\!
\frac{1}{2^{p}}\sum_{\blambda'}\left[\hat{P}_{i}^{\mr{s}}
\left(B_{\mr{s}}(k;\blambda'\cdot\blambda,
\hat{m}_{0}+\hat{m}\la'),F_{\mu}\blambda'\right)-
\hat{P}_{i}^{\mr{s}}
\left(B_{\mr{s}}(k;\blambda'\cdot\blambda,
\hat{m}_{0}+\hat{m}\la'),\blambda'\right)\right]
\\[4mm]
\mr{par:}\hspace{6mm}
\Delta^{\mr{p}}_{i+1}\!\!&\!\!=\!\!&\!\!\hat{P}_{i+1}^{\mr{p}}
(k,F_{\mu}\blambda)-\hat{P}_{i+1}^{\mr{p}}(k,\blambda)
\\[2mm]
\!\!&\!\!=\!\!&\!\!
\frac{1}{2^{2p}}\sum_{\blambda'',\bxi'}
\left[\hat{P}^{\mr{p}}_{i}\left(B_{\mr{p}}(k;{\blambda}''\cdot\bxi',
\bxi'\cdot\blambda,m_{0}+m\xi',
\hat{m}_{0}+\hat{m}\lambda''),F_{\mu}\blambda'\right)-\right.
\\[2mm]
\!\!&\!\!\!\!&\!\!
\hspace{40mm}
\left.\hat{P}^{\mr{p}}_{i}\left(B_{\mr{p}}(k;\blambda''\cdot\bxi',
\bxi'\cdot\blambda,m_{0}+m\xi',
\hat{m}_{0}+\hat{m}\lambda''),\blambda'\right)
\right]
\end{array}
\]
If at step $i$ the identities $\Delta_i^{\rm s}=\Delta_{i}^{\rm
p}=0$ are true, then
$\hat{P}_i(B(k),F_\mu\blambda)=
\hat{P}_i(B(k),\blambda)$  (for both types of dynamics)  and thus
$\Delta^{\mr{s}}_{i+1}=\Delta^{\mr{p}}_{i+1}=0$. Upon choosing
suitable initial conditions we can thus construct equilibrium
integrated densities with the stated property; combination with the
assumed ergodicity of the process then implies that the unique
solution must have the property. This completes the proof. We can
consequently write the stationary integrated densities in the form:
$\hat{P}^{\mr{s}}_{\infty}(k,\la)$ and
$\hat{P}^{\mr{p}}_{\infty}(k,\la)$, where $\lambda\in\{-1,1\}$
corresponds to the condensed pattern component:
\begin{eqnarray}
\mr{sequential:}\hspace{6mm}
\hat{P}_{i+1}^{\mr{s}}(k,\lambda)
\!&= &\!
\frac{1}{2}\sum_{\lambda'=\pm1}\hat{P}_{i}^{\mr{s}}
\left(B_{\mr{s}}(k\,;\,\lambda'\lambda\,,\,\hat{m}_0+\hat{m}\lambda'),
\lambda'\right)  \label{eq:condensed-densities-seq}
\\
\mr{parallel:}\hspace{6mm}
\hat{P}_{i+1}^{\mr{p}}(k,\lambda)
\!& =&\!
\frac{1}{4}\!\sum_{\lambda'',\xi'=\pm1}\!
\hat{P}_{i}^{\mr{p}}\left(B_{\mr{p}}(k\,;\,\lambda''\xi'\,,\,
\xi'\lambda\,,m_0+m\xi' \,,\,\hat{m}_0+\hat{m}\lambda''),
\lambda''\right)  \label{eq:condensed-densities-par}
\end{eqnarray}
One can also exploit the fact that the pattern variables
$\{\blambda\}_{\mr{seq}}$ and $\{\blambda\}_{\mr{par}}$ appear in
inner-products only, to simplify the non-trivial integrated
expressions of equation
(\ref{eq:nontrivial_seq},\ref{eq:nontrivial_par}). Since the
argumentation will be qualitative we will forget about the details
of these expressions and we will denote the integrated logarithmic
expressions simply by
\[
\Phi_{\rm s}(\blambda'\cdot\blambda)=
 \log \left[e^{\beta J_{s}(\blambda'\cdot\blambda)}+
    e^{-\beta J_{s}(\blambda'\cdot\blambda)}+
    k'e^{2\beta i(\hat{m}_{0}+\hat{m}\lambda')}\right]
\]
\[
  \Phi_{\rm p}(\blambda''\cdot\bxi',\bxi'\cdot\blambda)
=\log\left[A_{(+,+)}(\blambda''\cdot\bxi',\bxi'\cdot\blambda,\xi'm)
+A_{(-,+)}(\blambda''\cdot\bxi',\bxi'\cdot\blambda,\xi'm)\
k''\ e^{2\beta i(\hat{m}_0+\hat{m}\lambda'')}\right]
\]
Upon using the gauge transformations:
$\la_{\mu}=\tau_{\mu}\la_{\mu}'$ for sequential dynamics and
$\la_{\mu}''=\tau_{\mu}\xi_{\mu}'$,
$\la_{\mu}=\eta_{\mu}\xi_{\mu}'$ for parallel dynamics, where
$\tau_{\mu}$ and $v_{\mu}$ are auxiliary Ising variables, the
non-trivial parts of expressions (\ref{eq:nontrivial_seq}) and
(\ref{eq:nontrivial_par}) --corresponding to the integrals over the
stochastic variables $\{k_j\}$ after the distribution factorisation
(\ref{eq:factor_seq}) and (\ref{eq:factor_par}) and the `pure
state' ansatz-- take the form:
\begin{eqnarray*}
\mr{sequential:}
 & &
\frac{1}{2^{p}}\sum_{\blambda'}\sum_{\blambda}\int\! dk'\,
P_{\infty}^{\mr{s}}(k',\la')\ \Phi_{\mr{s}}
(\blambda'\cdot\blambda\,;\,k',\hat{m}_{0}+\hat{m}\la')
\\
&&=
\frac{p-1}{2^p}\sum_{\lambda'}\sum_{\boldtau}\int\! dk'\,
P_{\infty}^{\mr{s}}(k',\la')\
\Phi_{\mr{s}}(\sum_{\mu=1}^{p}\!\tau_{\mu}\,;\,k',\hat{m}_{0}+\hat{m}\la')
\\
&&
=\frac{p-1}{2^p}\sum_{\la'=\pm1}\sum_{\mc{N}_{\tau}=1}^{p}\int\! dk'\, \left(\!\!
\begin{array}{l}
p \\ \mc{N}_{\tau}
\end{array}\!\!\right)
P_{\infty}^{\mr{s}}(k',\la')\
\Phi_{\mr{s}}(2\mc{N}_{\tau}-p\,;\,k',\hat{m}_0+\hat{m}\la')
\\
\mr{parallel:}& &
\frac{1}{2^{2p}}\sum_{\blambda''}\sum_{\bxi'}\sum_{\blambda}\int\! dk''\
P_{\infty}^{\mr{p}}(k'',\lambda'')\
\Phi_{\mr{p}}(\blambda''\cdot\bxi',\bxi'\cdot\blambda, m\xi'\,;\,k'',\hat{m}_0+\hat{m}\lambda'')
\\
&&=
\frac{p-1}{2^{2p}}\sum_{\boldtau}\sum_{\xi'}\sum_{\boldeta}
\int\! dk''\
P^{\mr{p}}_{\infty}(k'',\tau_1\xi')\
\Phi_{\mr{p}}(\tau_1+\sum_{\mu=2}^{p}\tau_{\mu},
  \sum_{\mu=1}^{p}\eta_{\mu}, m\xi'\,;\,k'',
  \hat{m}_0+\hat{m}\tau_1\xi')
\\
& &=
\frac{(p-1)^2}{2^{2p} }
\sum_{\tau_1=\pm1}\sum_{\xi'=\pm1}\sum_{\mc{N}_{\tau}=2}^{p}
\sum_{\mc{N}_{\eta}=1}^{p}\int\! dk''\
\left(\!\!\begin{array}{l}
p-1 \\ \mc{N}_{\tau}
\end{array}\!\!\right)
\!
\left(\!\!\begin{array}{l}
p \\ \mc{N}_{\eta}
\end{array}\!\!\right)
P^{\mr{p}}_{\infty}(k'',\tau_1\xi')\
\times
\\
& &
\hspace{30mm}\times\
\Phi_{\mr{p}}(\tau_1+2\mc{N}_{\tau}-p+1,2\mc{N}_{\eta}-p,
m\xi'\,;\,k'',\hat{m}_0+\hat{m}\tau_1\xi')
\end{eqnarray*}
where $\mc{N}_{\tau}$ and $\mc{N}_{\eta}$ represent the number of
neurons with states equal to $+1$ in the configurations of
$\boldtau$, $\boldeta$ respectively. We have thus replaced all
summations over the $2^{p}$ configurations of the vectors
$\{\blambda,\blambda'\}_{\mr{seq}}$ and
$\{\blambda'',\bxi',\blambda\}_{\mr{par}}$ by summations over
binary- and $p$-state variables.


\subsection*{Phase Diagrams}

In order to calculate phase transitions and draw phase diagrams we
will first calculate the free energy surfaces
(\ref{eq:phis_seq}-\ref{eq:phis_par}), which at this stage are
still functions of the order parameters $m$ and $\hat{m}$. For
simplicity we will now set $\theta=0$. The distributions
$P_\infty^{\rm s}(\ldots)$ and $P_\infty^{\rm p}(\ldots)$ can be
calculated numerically via iteration of
(\ref{eq:condensed-densities-seq}-\ref{eq:condensed-densities-par}),
and bifurcations of the non-trivial values for the pure state
overlap from the trivial solution (if they exist) will then be
given as the solutions of the following fixed point problems:
\begin{equation}
\begin{array}{rllccl}
{\rm sequential:} & m=\partial_{i\hat{m}}F_{\rm seq}(i\hat{m})
& &{\rm at} & & i\hat{m}=-J_{\ell}^{(2)}m
\\[3mm]
{\rm parallel:} & m=\partial_{i\hat{m}}F_{\rm par}(i\hat{m},m) &
&{\rm at} & &i\hat{m}=\partial_{m}F_{\rm par}(i\hat{m},m)
\end{array} \label{eq:fixed_modelII}
\end{equation}
with
\begin{eqnarray}
F_{\rm seq}(i\hat{m})&=&
   -\frac{p-1}{2^p \beta}
   \sum_{\lambda'\pm1}\sum_{\mc{N}_\tau=1}^{p}\int dk'
   \left(\!\!\begin{array}{c}
   p \\ \mc{N}_{\tau}
   \end{array}\!\!\right)
   P_{\infty}^{\rm s}(k',\lambda')\
   \times
     \nonumber
 \\
& &
 \times
\log\left[e^{\beta[J_{s}^{(1)}+ J_{s}^{(2)}(2\mc{N}_\tau-p)]}
   +e^{-\beta[J_{s}^{(1)}+ J_{s}^{(2)}(2\mc{N}_\tau-p)]}\ k'
    e^{2\beta i\hat{m}\lambda'}\right]  \label{eq:mseq}
\\
F_{\rm par}(i\hat{m},m)& =&
   -\frac{(p-1)^2}{2^{2p} \beta}
   \sum_{\tau_1=\pm1}\sum_{\xi'=\pm1}
   \sum_{\mc{N}_{\tau}=2}^{p}\sum_{\mc{N}_{\eta}=1}^{p}
   \int dk''
   \left(\!\!\begin{array}{c}
   p-1 \\ \mc{N}_{\tau}
   \end{array}\!\!\right)
   \left(\!\!\begin{array}{c}
   p \\ \mc{N}_{\eta}
   \end{array}\!\!\right)
   P_{\infty}^{\rm p}(k'',\tau_{1}\xi')\times
   \nonumber
\\
&&   \nonumber
   \times\log\left[2\cosh[\beta J_{\ell}^{(2)}m\xi'+
   2\beta J_{s}^{(1)}+\beta J_{s}^{(2)}(\tau_1-2p+2\mc{N}_{\tau}+2\mc{N}_{\eta}+1)]+\right.
\\
&&
\left.  +2\cosh[\beta J_{\ell}^{(2)}m\xi'+
         \beta J_{s}^{(2)}[2\mc{N}_{\eta}-2\mc{N}_{\tau}+\tau_1+1]]\
          k''\ e^{2\beta i\hat{m}\tau_1\xi'}
          \right]      \nonumber
\\
&&
-\frac{p-1}{2^{2p}\beta}
\sum_{\xi=\pm1}\sum_{\mc{N}=1}^{p}
\left(\begin{array}{c}
p \\ \mc{N}
\end{array}\right)
\log \left[\frac{\cosh[\beta J_\ell^{(2)}m\xi
+J_s^{(1)}+J_{s}^{(2)}(2\mc{N}-p)]}
{\cosh[\beta J_\ell^{(2)}m\xi
-J_s^{(1)}-J_{s}^{(2)}(2\mc{N}-p)]}\right]
          \label{eq:mpar}
\end{eqnarray}
As in the solution of model (\ref{eq:model_Ib}) in section
\ref{sec:sol_modelI}, the parallel dynamics fixed-point problem
(\ref{eq:fixed_modelII},\ref{eq:mpar}) takes the form of a set of
coupled equations, which makes the evaluation of bifurcation points
essentially more laborious.

\begin{figure}[t]
\vspace*{62mm}
\hbox to
\hsize{\hspace*{-1mm}\includegraphics{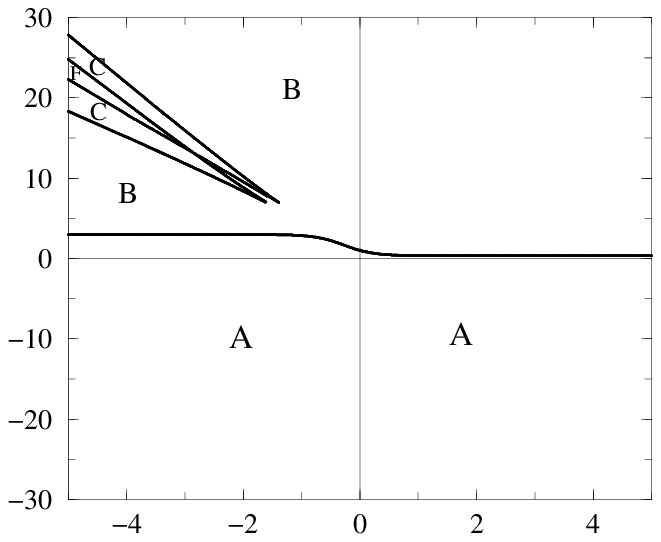}\hspace*{1mm}}
\vspace*{-5mm}
\hbox to
\hsize{\hspace*{76mm}\includegraphics{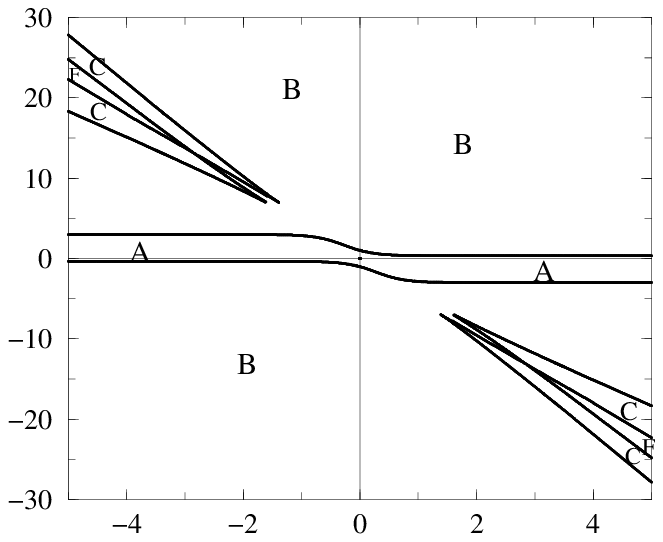}\hspace*{-76mm}}
\vspace*{-32mm}
\begin{picture}(200,85)(10,50)
\hspace*{-5mm}
{\Large
\put(10,145){ $\beta J_{\ell}^{(2)}$}
\put(230,145){ $\beta J_\ell^{(2)}$}
\put(140,50){ $\beta J_{s}^{(2)}$}
\put(350,50){ $\beta J_{s}^{(2)}$}
}
\end{picture}
\vspace*{2mm}
\caption{\small  Phase diagrams of model (\ref{eq:model_2})
for $p=2$ and with $J_s^{(1)}=J_{\ell}^{(1)}=0$, for sequential
(left picture) and parallel dynamics (right picture). Lines
separate regions with different numbers of locally stable solutions
for the pure state overlap $m$, calculated from equations
(\ref{eq:mseq}-\ref{eq:mpar}). In region {\textbf{A}} the only
stable solution is the trivial one. The transition
{\textbf{A}}$\to${\textbf{B}} is second-order, whereas
{\textbf{B}}$\to${\textbf{C}} and {\textbf{F}}$\to${\textbf{C}} are
first-order (see also figure \ref{fig:bifurcation_p2}). In regions
{\textbf{B}},{\textbf{C}} and {\textbf{F}} there are 1,2 and 3
locally stable $m>0$ states, respectively (see also free energy
graphs, figure \ref{fig:simulations}).}
\label{fig:diagram_p2}
\end{figure}

The solutions of the above equations, for
$J_{s}^{(1)}=J_{\ell}^{(1)}=0$ and for $p=2$, are shown in the
phase diagrams of figure \ref{fig:diagram_p2}. One distinguishes
between four different regions, dependent on the number of locally
stable `pure state' solutions: region {\textbf{A}} with $m=0$ only,
region {\textbf{B}} with one locally stable $m>0$ state (and one
$m<0$), regions {\textbf{C}} with two locally stable $m>0$ states
(and two $m<0$ ones) and region {\textbf{F}} with three $m>0$
states (and three $m<0$ ones). Note that region {\textbf{F}} is
created at the point where regions {\textbf{C}} start
overlapping. The transition {\textbf{A}}$\to${\textbf{B}} is
second-order, whereas {\textbf{B}}$\to${\textbf{C} and
{\textbf{F}}$\to${\textbf{C}} are first-order. The two
qualitatively different types of bifurcations are also shown in
figure \ref{fig:bifurcation_p2} (left picture), 
where we draw the solution(s) of the
overlap $m$ as a function of $\beta J_{\ell}$ along the line $\beta
J_s=-1.8$ (a line crossing regions {\textbf{A}},{\textbf{B}} and
{\textbf{C}}). The zero noise region ($T=\beta^{-1}=0$) for the 
phase diagram of figure 
\ref{fig:diagram_p2} is shown in figure \ref{fig:bifurcation_p2} 
(right picture)
where we draw the transition lines separating recall regimes {\textbf A}
,{\textbf B}, {\textbf C} and {\textbf F} 
in the $(J_{\ell}^{(2)},T)$ plane for $J_{s}^{(2)}=-4$.

In equations (\ref{eq:mseq}) and (\ref{eq:mpar}) we observe that,
due to the explicit appearance of the variable $p$ in the solution
of the overlap order parameter (which is due to short-range
interactions, originating from expressions of the form
$\exp[\beta(J_s^{(1)}+J_s^{(2)}\sum_{\mu=1}^{p}\xi^\mu_i\xi^\mu_{i+1})]$,
see e.g.\@ (\ref{eq:Rseq})), it will no longer be true that the
pure state ansatz leads to solutions which are independent of the
number of stored patterns, as is the case for standard mean-field
Hopfield networks. This is also shown in the phase diagrams of
figure \ref{fig:diagram_p15}, which have been constructed from
(\ref{eq:mseq},\ref{eq:mpar}) with $J_{s}^{(1)}=J_{\ell}^{(1)}=0$
and for $p=15$. We observe a significant increase in the number of
transition lines, as well as additional transition lines appearing
in the quadrant $J_{\ell}^{(2)},J_s^{(2)}>0$ (and also in
$J_{\ell}^{(2)},J_{s}^{(2)}<0$ for the parallel case). Such effects
become more and more prominent as the number of patterns increases.
It is also worth noting that figures \ref{fig:diagram_p2} and
\ref{fig:diagram_p15} imply that all `pairs' of first-order
transition lines point to the origin of the
$\{J_s^{(2)},J_{\ell}^{(2)}\}$ plane.

Finally, due to the occurrence of imaginary saddle-points in
(\ref{eq:fixed_modelII}) and our strategy to eliminate the variable
$\mh$ by using the equation $\partial_m\phi(m,\mh)=0$, it need not
be true that the saddle-point with the lowest value of
$\phi(m,\mh)$ is the minimum of $\phi$ (complex conjugation can
induce curvature sign changes, and in addition the minimum could
occur at boundaries or as special limits). To remove this
uncertainty we have evaluated the sequential dynamics free energy
$\phi_{\rm seq}(m,\hat{m})$ (\ref{eq:phis_seq}) after elimination
of the conjugate variable $\hat{m}$ by using
$\partial_{m}\phi(m,\hat{m})=0$ (middle row of figure
\ref{fig:simulations}) and $\partial_{\hat{m}}\phi(m,\hat{m})=0$
(lower row of figure \ref{fig:simulations}) which shows that,
although the convexity of the free energy graphs is indeed
affected, the location of the minima is not.
\vspace*{3mm}

\begin{figure}[t]
\vspace*{62mm}
\hbox to
\hsize{\hspace*{-1mm}\includegraphics{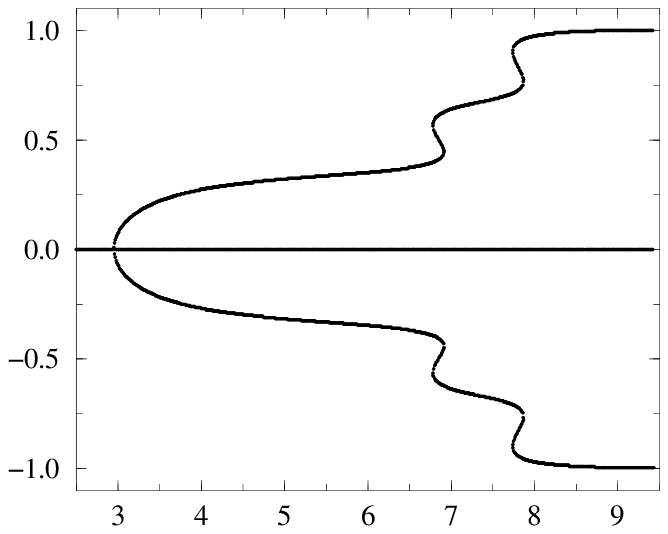}\hspace*{1mm}}
\vspace*{-5mm}
\hbox to
\hsize{\hspace*{76mm}\includegraphics{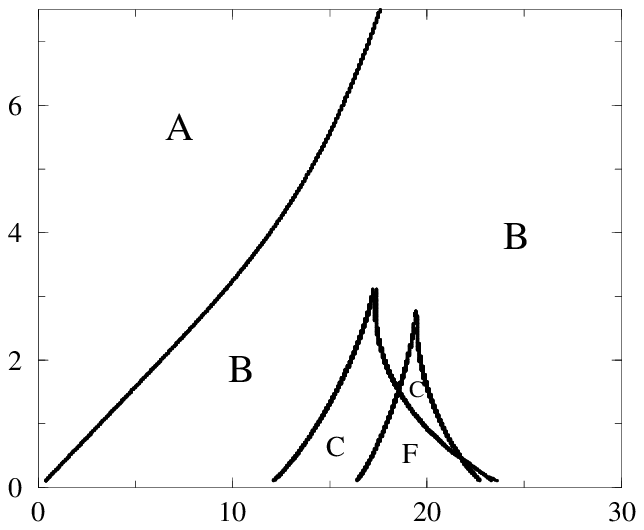}\hspace*{-76mm}}
\vspace*{-32mm}
\begin{picture}(200,85)(10,50)
\hspace*{-5mm}
{\Large
\put(10,145){ $m$}
\put(240,145){ $T$}
\put(140,50){ $\beta J_{\ell}^{(2)}$}
\put(350,50){ $ J_{\ell}^{(2)}$}
}
\end{picture}
\vspace*{2mm}
\caption{\small Left: Sequential dynamics bifurcation diagram
corresponding to the phase diagram of figure \ref{fig:diagram_p2}
along the line $\beta J_s^{(2)}=-1.8$. The transition
{\textbf{A}}$\to${\textbf{B}} in figure \ref{fig:diagram_p2} is
shown here as a continuous bifurcation of the trivial solution
whereas the two other bifurcations (at $\beta J_{\ell}^{(2)}\approx
6.76$ and $\beta J_{\ell}^{(2)}\approx 7.7$) correspond to the
first-order transitions {\textbf{B}}$\to${\textbf{C}}. Right:
Alternative presentation of the phase diagram of figure
\ref{fig:diagram_p2} drawn in the $(J_{\ell}^{(2)},T)$ plane (with 
$T=\beta^{-1}$)  for $J_s^{(2)}=-4$.}

\label{fig:bifurcation_p2}
\end{figure}

\begin{figure}[t]
\vspace*{62mm}
\hbox to
\hsize{\hspace*{-1mm}\includegraphics{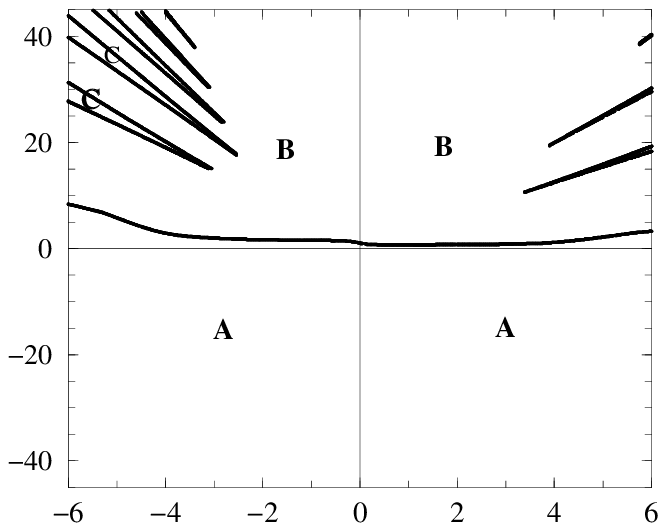}\hspace*{1mm}}
\vspace*{-5mm}
\hbox to
\hsize{\hspace*{75mm}\includegraphics{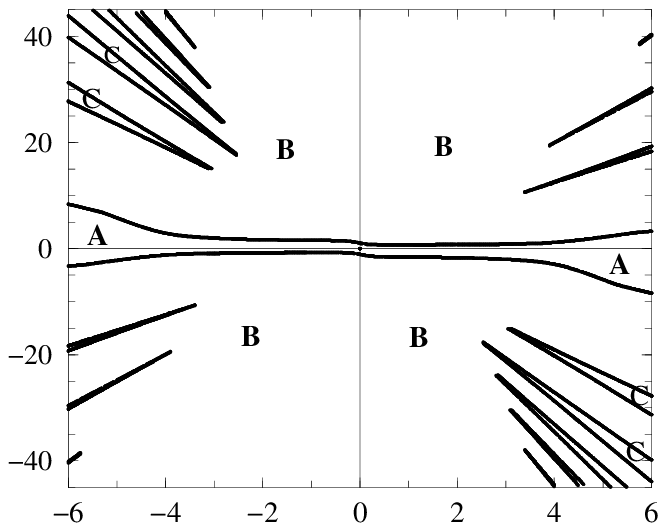}\hspace*{-75mm}}
\vspace*{-32mm}
\begin{picture}(200,85)(10,50)
\hspace*{-5mm}
{\Large
\put(10,145){ $\beta J_{\ell}^{(2)}$}
\put(230,145){ $\beta J_{\ell}^{(2)}$}
\put(140,50){ $\beta J_{s}^{(2)}$}
\put(350,50){ $\beta J_{s}^{(2)}$}
}
\end{picture}
\vspace*{2mm}
\caption{\small  Phase diagrams of model (\ref{eq:model_2})
for $p=15$ and with $J_{\ell}^{(1)}=J_s^{(1)}=0$, for sequential
dynamics (left picture) and parallel dynamics (right picture).
Lines separate regions with different numbers of locally stable
solutions of the pure state overlap $m$, calculated from equations
(\ref{eq:mseq}-\ref{eq:mpar}). Compared with the cases $p=1$
(figure \ref{fig:diagrams}) and $p=2$ (figure
\ref{fig:diagram_p2}), we observe a significant increase in the
number of transition lines, caused by the explicit dependence of
equations (\ref{eq:mseq}-\ref{eq:mpar}) on $p$. The diagrams
involve three regions: region {\textbf{A}} where $m=0$ only, region
{\textbf{B}} with one $m>0$ locally stable state and regions
{\textbf{C}} (appearing inside each of the transition-line pairs)
with two locally stable  $m>0$ states. The transition
{\textbf{A}}$\to${\textbf{B}} is second-order, whereas all
transitions {\textbf{B}}$\to${\textbf{C}} are first-order. Also
note the appearance of further transition lines in the upper right
quadrant, where $J_{\ell}^{(2)},J_{s}^{(2)}>0$ (and the lower left
quadrant for the parallel dynamics case).}
\label{fig:diagram_p15}
\end{figure}

\section{Benchmark Tests}

We now compare our results with simple benchmark cases.
First, the solution of model (\ref{eq:model_2}) should reduce to
the model of Amit et al \cite{amitetal1}, for regimes where $p\ll
N$, upon removing short-range connectivity, i.e.\@ for
$J_{s}^{(1)}=J_{s}^{(2)}=0$. Indeed we find that in this limit the
probability distributions $P^{\rm seq}_{\infty}(k)$ and
$P_\infty^{\rm par}(k)$ of the stochastic variables
(\ref{eq:seqmap}) and (\ref{eq:parmap}) both reduce to the delta
peak: $\delta[k-1]$. This simplifies the solution of our problem
and allows us to write for the free energies:
\begin{eqnarray*}
\phi_{\rm seq}(m,\hat{m}) & = &
-im_0\hat{m}_0-im\hat{m}-m_0\theta-\frac12J_{\ell}^{(1)}m_0^2+\frac12 J_{\ell}^{(2)}m^2
-\frac1\beta \bra \log 2\cosh[\beta
i(\hat{m}_0+\hat{m}\xi)]\ket_\xi
\\
\phi_{\rm par}(m,\hat{m}) & = &
-im_0\hat{m}_0-im\hat{m}-m_0\theta-
\\
& &
-\frac1\beta\bra\log 2\cosh[\beta i(\hat{m}_0+\hat{m}\xi)]\ket_\xi
-\frac1\beta\bra \log
2\cosh[\beta(\theta+J_{\ell}^{(1)}m_0+J_{\ell}^{(2)}m\xi)]\ket_{\xi}
\end{eqnarray*}
Simple differentiation with respect to
$\{m_0,\hat{m}_0,m,\hat{m}\}$ verifies that in the mean-field limit
the pure order parameter solutions reduce to
$m=\bra\xi\,\tanh[\beta(\theta+J_{\ell}^{(1)}m_0+\xi
J_{\ell}^{(2)}m)]\ket_{\xi}$ and
$m_0=\bra\tanh[\beta(\theta+J_{\ell}^{(1)}m_0+\xi
J_{\ell}^{(2)}m)]\ket_{\xi}$ as they should.

Our second benchmark test is provided by the exact solution of
model (\ref{eq:model_Ib}), section \ref{sec:sol_modelI}. We can
immediately map model (\ref{eq:model_2}) to model
(\ref{eq:model_Ib}) by setting $J_\ell^{(2)}=J_s^{(2)}=0$. We then
find that the key variables $\{k_j\}$ of equations
(\ref{eq:kseq}-\ref{eq:kpar}) are given by a simple deterministic
map. In fact, for both sequential and parallel dynamics we find
that $k_{\rm seq}$ and $k_{\rm par}$ evolve towards the same fixed
point:
\[
k=e^{2\beta J_s}e^{-\beta i\hat{m}}\left[-\sinh[\beta i\hat{m}]
+\sqrt{\sinh^{2}[\beta i\hat{m}]+e^{-4\beta J_s}}\right]
\]
which (at the relevant saddle points) can be verified to lead to
the fixed-point equation (\ref{eq:saddle}) of model
(\ref{eq:model_Ib}), namely
\[
m=G(m;J_\ell,J_s)
\hspace{10mm}
{\rm with }
\hspace{10mm}
G(m;J_\ell,J_s)=\frac{\sinh[\beta J_\ell m]}
{\sqrt{\sinh^2[\beta J_{\ell} m]+e^{-4\beta J_s}}}
\]

Thirdly, we have also compared the free energies of model
(\ref{eq:model_2}), as given by our present solution, to that which
one finds when using the alternative random-field technique of
\cite{rujan}. The latter relies on performing the spin summations
in $R=\sum_{\bs}\mc{F}(\bs)$ (\ref{eq:Rseq}) and deriving
appropriate functions $A(\bxi\cdot\bxi')$ and $B(\bxi\cdot\bxi')$
such that the identity
$\cosh[\beta(J_s(\bxi\cdot\bxi')\s'-i\hat{m}\xi)]=
\exp[\beta(A(\bxi\cdot\bxi')\s'+B(\bxi\cdot\bxi'))]$
is true for $\s'\in\{-1,1\}$. For instance, for the expression
(\ref{eq:Rseq}) of sequential dynamics this leads to:
\[
-\lim_{N\to\infty}\frac{1}{\beta N}\log R_{\rm seq}
=
-\lim_{N\to\infty}\frac{1}{2\beta N}\sum_{i=1}^N
    \log\left[4\cosh[\beta(J^s_{i,i+1}+h_{i})]
     \,\cosh[\beta(J^s_{i,i+1}-h_{i})]\right]
\]
\[
{\rm where}
\hspace{10mm}
h_{i+1}=i\hat{m}\xi_{i+1}-\frac12 \log\left[\frac{\cosh[\beta(J^s_{i,i+1}-h_{i})]}
         {\cosh[\beta(J^s_{i,i+1}+h_{i})]}\right]
\hspace{10mm}
J^s_{l k}=J_s^{(1)}+J_s^{(2)}\bxi_l\cdot\bxi_k
\]
(with $h_1=i \hat{m}\xi_1$) and is in complete agreement with the
free energy as found from
(\ref{eq:phis_seq},\ref{eq:nontrivial_seq},\ref{eq:integrated_seq}).

Finally, for the special case
$J_{s}^{(1)}=J_{\ell}^{(1)}=J_{\ell}^{(2)}=0$ and $p=1$
(short-range bond disorder and absence of long-range interactions)
our model reduces to the classical short-range random-bond Ising
model \cite{brandtgross}, in which we expect the integrated density
$\hat{P}_{\!\infty}(k)$ (\ref{eq:condensed-densities-seq}) to
acquire, at least in certain parameter regions, the form of the
highly non-analytic Devil's Staircase \cite{bruinsma,aeppli}. In
this special case the density (\ref{eq:condensed-densities-seq})
reduces (at saddle points $\{i\hat{m}_0=-\theta,i\hat{m}=0\}$) to
\[
\hat{P}_{i+1}(k,\lambda)=\frac12
  \left\{\hat{P}_i\left(e^{2\beta\theta}\ \frac
  {ke^{\beta J_s\lambda}-e^{-\beta J_s\lambda}}
  {e^{\beta J_s\lambda}-k e^{-\beta J_s\lambda}}\ ,1\right)+
  \hat{P}_{i}\left(e^{2\beta\theta}\ \frac
  {ke^{-\beta J_s\lambda}-e^{\beta J_s\lambda}}{e^{-\beta J_s\lambda}-
  k e^{\beta J_s\lambda}}\ ,-1\right)\right\}
\]
where $\lambda=\pm 1$ represents bond-disorder. One can now prove
by induction that if the identity
$\hat{P}_{i}(k,1)=\hat{P}_{i}(k,-1)$ is true then also
$\hat{P}_{i+1}(k,1)=\hat{P}_{i+1}(k,-1)$, so that (assuming
ergodicity and uniqueness of the stationary density) the above
expression reduces to a single recursive equation:
\[
\hat{P}_{i+1}(k)=\frac12
  \left\{\hat{P}_i\left(e^{2\beta\theta}\ \frac
  {ke^{\beta J_s}-e^{-\beta J_s}}
  {e^{\beta J_s}-k e^{-\beta J_s}}\right)+
  \hat{P}_{i}\left(e^{2\beta\theta}\ \frac
  {ke^{-\beta J_s}-e^{\beta J_s}}{e^{-\beta J_s}-
  k e^{\beta J_s}}\right)\right\}
\]
which is recognised as equation 10 of Ref.\@ \cite{brandtgross}, upon a simple
re-definition of our stochastic variables: $\{k_n=e^{-2\beta
i\hat{m}_0} R_{n,\da}/R_{n,\ua}\}\ra
\{e^{2\beta i\hat{m}_0} R_{n,\ua}/R_{n,\da}\}$
in (\ref{eq:define_kseq}). In the present benchmark case we have
also verified the identity found in \cite{skantzos} relating
short-range random-field models between sequential and parallel
dynamics, namely
\[
\psi_{\rm seq}\left(\psi_{\rm
seq}(k\,;\,\xi''\xi',\theta)\,,\xi'\xi,\theta\right)=
\psi_{\rm par}\left(k\,;\xi''\xi',\xi'\xi,\theta\right)
\]
(which is the key identity to prove that in the thermodynamic limit
sequential and parallel random-field models lead to the same
physical states). Here, $\psi_{\rm seq}(\ldots)$ and $\psi_{\rm
par}(\ldots)$ correspond to the functions defined in
(\ref{eq:seqmap}) and (\ref{eq:parmap}).

\section{Theory Vs Simulations \label{sec:simulations}}

\begin{figure}[t]
\vspace*{50mm}
\hbox to
\hsize{\hspace*{4mm}\includegraphics{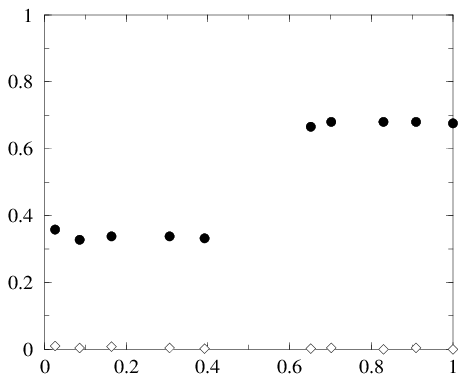}\hspace*{-4mm}}
\vspace*{-5mm}
\hbox to
\hsize{\hspace*{55mm}\includegraphics{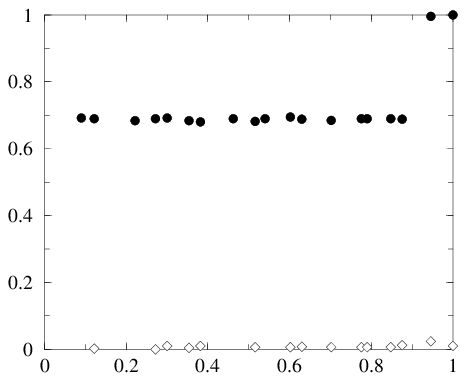}\hspace*{-55mm}}
\vspace*{-5mm}
\hbox to
\hsize{\hspace*{107mm}\includegraphics{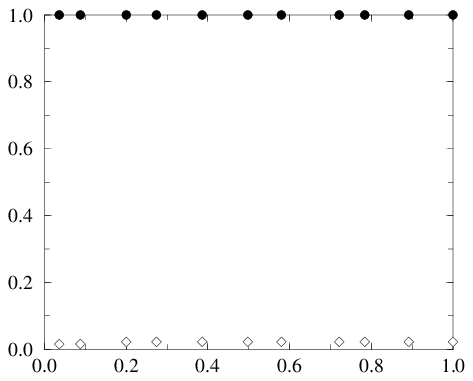}\hspace*{-107mm}}
\vspace*{-32mm}
\begin{picture}(200,85)(10,50)
{ \Large
\put(0,120){$m_{equil}$}
\put(100,50){$m_{init}$}
\put(240,50){$m_{init}$}
\put(380,50){$m_{init}$}
\put(80,180){Region \textbf{C}}
\put(225,180){Region \textbf{F}}
\put(370,180){Region \textbf{B}}
\hspace*{-5mm}
}
\end{picture}

\vspace*{50mm}
\hbox to
\hsize{\hspace*{4mm}\includegraphics{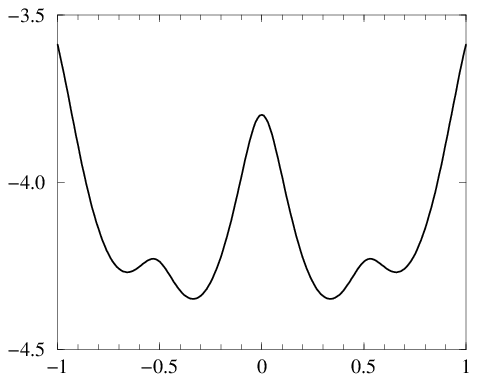}\hspace*{-4mm}}
\vspace*{-5mm}
\hbox to
\hsize{\hspace*{55mm}\includegraphics{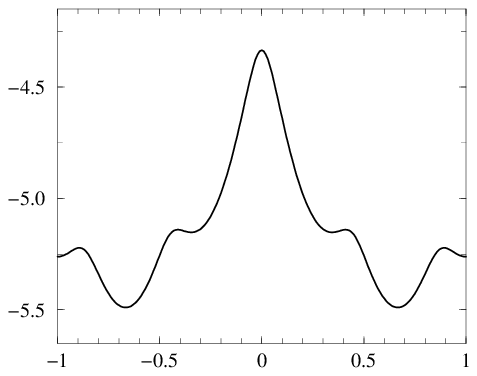}\hspace*{-55mm}}
\vspace*{-5mm}
\hbox to
\hsize{\hspace*{107mm}\includegraphics{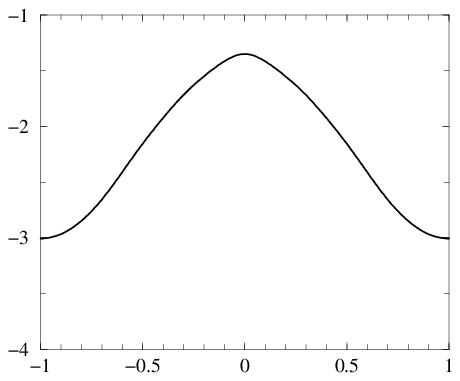}\hspace*{-107mm}}
\vspace*{-32mm}
\begin{picture}(200,85)(10,50)
{ \Large
\put(-5,120){$\phi(m)$}
\put(100,50){$m$}
\put(250,50){$m$}
\put(400,50){$m$}
\hspace*{-5mm}
}
\end{picture}

\vspace*{50mm}
\hbox to
\hsize{\hspace*{4mm}\includegraphics{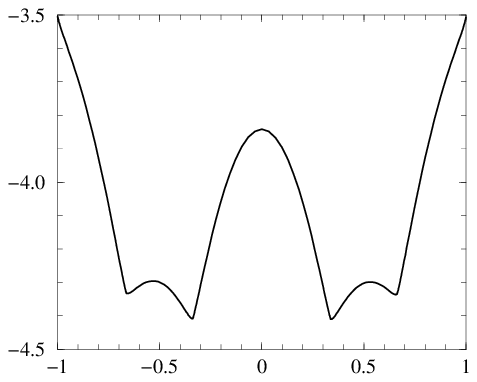}\hspace*{-4mm}}
\vspace*{-5mm}
\hbox to
\hsize{\hspace*{55mm}\includegraphics{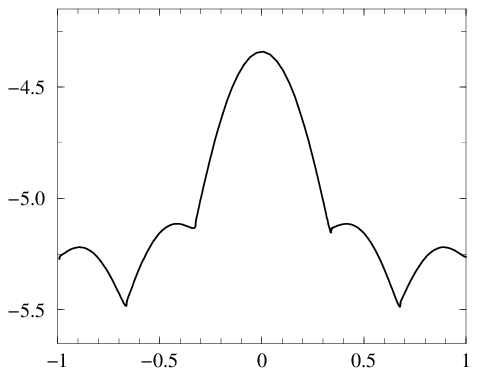}\hspace*{-55mm}}
\vspace*{-5mm}
\hbox to
\hsize{\hspace*{107mm}\includegraphics{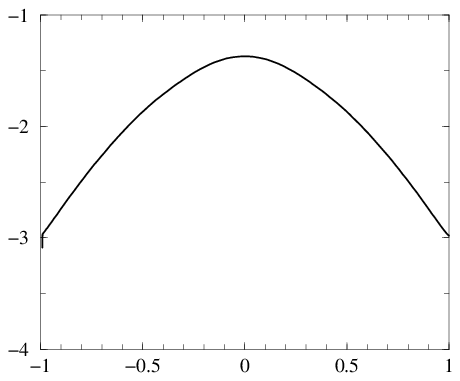}\hspace*{-107mm}}
\vspace*{-32mm}
\begin{picture}(200,85)(10,50)
{ \Large
\put(-5,120){$\phi(m)$}
\put(100,50){$m$}
\put(250,50){$m$}
\put(400,50){$m$}
\hspace*{-5mm}
}
\end{picture}
\vspace*{2mm}
\caption{\small Upper row: sequential dynamics simulation results
of the dynamical process (\ref{eq:dynamics}), with model
(\ref{eq:model_2}),  for a system with $p=2$ patterns. These were
carried out in three different regions {\textbf{C}},{\textbf{F}}
and {\textbf{B}} of the phase diagram of figure
\ref{fig:diagram_p2}. System size: $N=1,\!000$. Initial conditions
are random, subject to prescribed correlations with pattern
$\{\xi^1_i\}$. We show the equilibrium state $m(t\to\infty)$ of the
`pure state' overlap $m_1$ (full circles), as well as the overlap
$m_2$ (open diamonds), as functions of the initial state $m(t=0)$.
Finite size effects are of the order $\mc{O}(N^{-1/2})\approx
0.03$. Middle and lower rows: free energy per neuron $\phi_{\rm
seq}(m,i\hat{m})$, after elimination of the conjugate order
parameter $\hat{m}$ via $\partial_m\phi_{\rm seq}(m,i\hat{m})=0$
(middle row), and similarly after elimination of $\hat{m}$ via
$\partial_{\hat{m}}\phi_{\rm seq}(m,i\hat{m})=0$ (lower row). Left
column: $\beta J_{\ell}^{(2)}=14$ and $\beta J_s^{(2)}=-3.5$
(region {\textbf{C}} of the phase diagram of figure
\ref{fig:diagram_p2}), middle column: $\beta J_{\ell}^{(2)}=18.5$
and $\beta J_s^{(2)}=-4$ (region {\textbf{F}}), and right column:
$\beta J_{\ell}^{(2)}=8$ and $\beta J_s^{(2)}=-3.5$ (region
{\textbf{B}}). For all graphs $J_\ell^{(1)}=J_s^{(1)}=0$.}
\label{fig:simulations}
\end{figure}

In order to test our results further, we have performed extensive
simulation experiments of the process (\ref{eq:dynamics}), for
model (\ref{eq:model_2}). In all cases the initial state is
prepared randomly, with non-zero correlation only with pattern
$\{\xi_i^1\}$. Our simulation results for the model which gives the
phase diagram of figure \ref{fig:diagram_p2} ($p=2$ and
$J_{\ell}^{(1)}=J_s^{(1)}=0$) are shown in figure
\ref{fig:simulations} (upper row) where we
draw the equilibrium
value of the recall overlap $m_1(t\to\infty)$ as a function of the
initial state $m_1(t=0)$. We have performed our experiments for
three different regions of the phase diagram: region {\textbf{B}}
(one $m>0$  stable state), region {\textbf{C}} (two $m>0$ stable
states) and region {\textbf{F}} (three $m>0$ stable states). In
regions {\textbf{B}} and {\textbf{C}} the simulation experiments
verify the appearance and location of multiple ergodic sectors; to
compare with the theoretical results, see also the free
energy graphs in the middle and lower pictures of figure
\ref{fig:simulations}. In region {\textbf{F}} the simulation
experiments show that the system can enter only two possible
domains of attraction (excluding thus the theoretically predicted
state $m\approx 0.32$). This is due to ($i$) the system's finite
size ($N=1,\!000$), in combination with ($ii$) the (relatively)
small energy barrier separating the two physical states $m_a\approx
0.32$ and $m_b\approx 0.68$  (this allows the system to move from
state $m_a$ to $m_b$  with a non-negligible probability). Our
restriction to system size $N=1,\!000$ is prompted  by the
extremely long equilibration times (of the order of $\mc{O}(10^6)$
flips/spin). This, in turn, is due to domain formation: large
clusters of neurons tend to freeze in specific configurations. As a
consequence, in order for neurons to flip, the entire domain has to
flip. In figure \ref{fig:dynamics} (left graph) we show the value
of the condensed overlap as a function of time in region
{\textbf{F}}. We see that, starting from an initial state
$m_1(0)\approx 0.09$, the system gradually approaches the
theoretically predicted locally stable state, where it indeed stays
for a period of $\approx 4\cdot 10^5$ flips/spin. Due to finite
size effects, however, this state is thermodynamically unstable. A
sudden transition to a new meta-stable state is then observed,
generated by the flipping of entire domains. Equilibrium is reached
in these simulations at about $2\cdot 10^6$ flips/spin, where a
second and final jump transition takes place. In the right graph of
figure \ref{fig:dynamics} we show a simulation experiment carried
out in region {\textbf{C}}, starting from initial conditions
$m_1(0)\approx 0.08$. Here equilibrium is reached after about
$10^6$ flips/spin, and due to a (relatively) high energy barrier
separating the two $m>0$ physical states (see energy graphs in
regions {\textbf{C}}, left column of figure \ref{fig:simulations})
there is no domain-related transition. In all our experiments the
value of the non-selected pattern overlap $m_2(t)$ is found to
remain zero (open diamond points), which justifies {\em a
posteriori} our pure state ansatz.

\section{Discussion}

In this paper we have presented an exact equilibrium solution for a
specific class of spatially structured Ising spin (attractor)
neural network models, in which there is competition induced by the
presence of two qualitatively different types  of synaptic
interactions: those operating only between nearest neighbours in a
1D chain (short-range), and those operating between any pair of
neurons (long-range). The values taken by the interactions present
are given by Hebbian-type rules, as in the more familiar mean-field
attractor networks. We have solved these models by using a
combination of mean- and random-field techniques  for both
sequential and parallel dynamics. As in the standard 1D RFIM-type
models our expressions for the disorder-averaged free energy per
neuron take the form of integrals over the distribution of a random
variable, which represents the ratio of conditioned partition
functions. This distribution can be 
\clearpage
\begin{figure}[t]
\vspace*{60mm}
\hbox to
\hsize{\hspace*{-1mm}\includegraphics{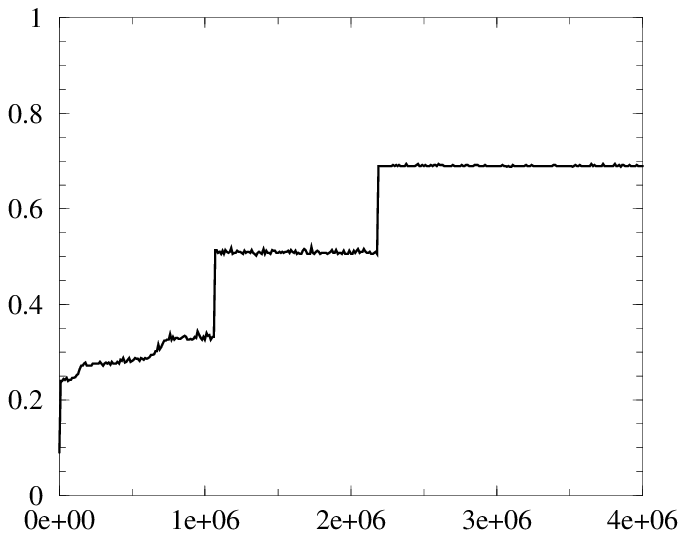}\hspace*{1mm}}
\vspace*{-5mm}
\hbox to
\hsize{\hspace*{77mm}\includegraphics{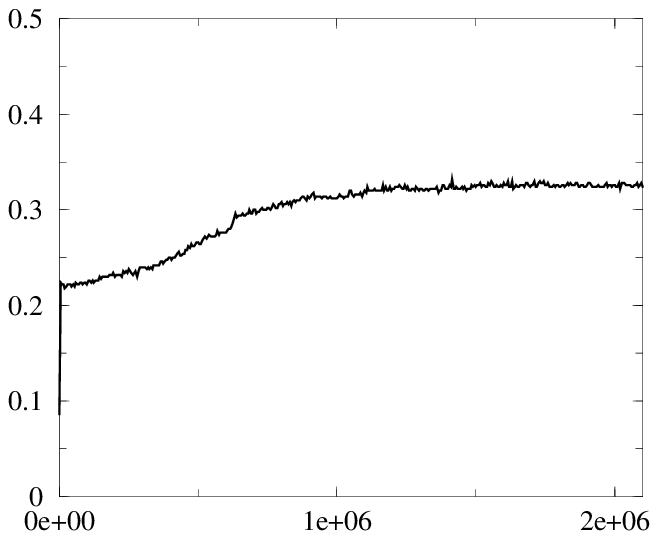}\hspace*{-77mm}}
\vspace*{-32mm}
\begin{picture}(200,85)(10,50)
{
\hspace*{-5mm}
\put(10,140){ \Large $m(t)$}
\put(90,50){ {\Large$t$}\ {\small$(flips/spin)$}}
\put(230,140){ \Large $m(t)$}
\put(320,50) { {\Large$t$}\ {\small$(flips/spin)$}}
}
\end{picture}
\vspace*{1mm}
\caption{\small  Simulation results for regions {\textbf{F}} (left
picture) and {\textbf{C}} (right picture) of the phase diagram of
figure \ref{fig:diagram_p2}. We show the evolution of the `pure
state' overlap order parameter as a function of time. In region
{\textbf{F}} (left) the theory predicts a locally stable state at
$m\approx 0.32$, which, due to finite size effects, appears here
only as a meta-stable state. Two prominent jump transitions occur,
until finally equilibrium is reached, at $m\approx 0.68$ (the jumps
indicate domain-flipping). In region {\textbf{C}} (right) full
equilibration still requires simulation times of the order of
$\mc{O}(10^6)$ flips/spin, after which the result is in true
agreement with the theory. }
\label{fig:dynamics}
\end{figure}
\noindent
evaluated numerically without
much effort, and the key macroscopic observables then follow via
simple differentiation.

We found that there are regions in parameter space where
information processing between the two types of synaptic
interactions can induce phenomena which are quite novel in the
arena of associative memory models, such as the appearance of
multiple locally stable states, and of first-order transitions
between them, even for finite $p$ and upon making the `pure state'
ansatz. These peculiarities come to life particularly in regions
where short- and long-range synapses compete most strongly, for
instance, where one has Hebbian long-range interactions in
combination with anti-Hebbian  short-range ones, and they become
more evident when increasing the number of stored patterns.
Particularly in the upper left quadrant of parameter space
$\{J_{\ell}>0, J_s<0\}$ one observes the appearance of an
increasing number of dynamic transition lines (first- and
second-order ones). This feature is in sharp contrast with the
conventional (infinite-range) Hopfield-like networks \cite{amitetal1},
where for finite $p$ the `pure state' ansatz automatically renders
the remaining order parameter independent of the number of
patterns stored. Phenomena such as simultaneous existence of
multiple locally stable states (in which the quality of pattern
recall depends crucially on initial conditions) can play a
potentially useful role: poor cue signals can no longer evoke
pattern recall. We also found that parallel dynamics transition
lines in parameter space are exact reflection in the origin of
those in sequential dynamics and that the relevant macroscopic
observables can be obtained from those of sequential dynamics via
simple transformations.  Simulation experiments also show that the
dynamics of the model are highly non-trivial, with plateaus and
jump discontinuities, caused by complex domain formation and domain
interaction, which would justify a study in itself.

In a similar fashion one could now also study more complicated
systems, where (in addition to the long-range synapses) the
short-range synapses reach beyond nearest neighbours. Such models
can still be solved using the techniques employed here. A different
type of generalisation would be to allow for a type of competition
between synapses  which would correspond to having stored patterns
with different (pattern dependent) embedding strengths, as in
\cite{viana}. All these will be subjects of a future work.

\subsection*{Acknowledgments}
It is our pleasure to thank Jort van Mourik for useful suggestions on
the simulations.

\end{document}